\documentclass[12pt]{iopart}

\usepackage{amssymb}
\usepackage[english]{babel}

\usepackage{epsfig}

\begin{document}
\baselineskip = 20pt

\def\gvect#1{\mbox{\boldmath $#1$}}

\newcommand{\TIT}[1]{\begin{center}\shadowbox{\Huge\sc \vbox{#1}}\end{center}}
\newcommand{\heading}[1]{\begin{center}\shadowbox{\Large\bf \vbox{#1}}
\end{center}}

\tableofcontents
\newpage


\title{Critical Behaviour of Irreversible Reaction Systems}
\author{Ernesto Loscar and Ezequiel V. Albano}
\address{ Instituto de Investigaciones Fisicoqu\'{\i}micas Te\'oricas y Aplicadas 
(INIFTA) \\ 
Sucursal 4, Casilla de Correo 16. (1900) La Plata \\ Argentina}

\begin{abstract}
An introductory review on the critical behaviour of 
some irreversible reaction systems is given. The study of these 
systems has attracted great attention during
the last decades due to, on the one hand, the rich and complex 
underlying physics, and on the other hand, their relevance for 
numerous technological applications in heterogeneous catalysis, corrosion 
and coating, development of microelectronic devices, etc..
The review is focuses on recent advances 
in the understanding of irreversible phase transitions (IPT's) providing a 
survey of the theoretical development of the field during the last decade,
as well as a detailed discussion of relevant numerical simulations.
The Langevin formulation for the treatment of second-order IPT's is discussed.
Different Monte Carlo approaches are also presented in detail
and the finite-size scaling analysis of second-order IPT's is described.
Special attention is devoted to the description of recent progress in the
study of first-order IPT's observed upon catalytic oxidation of carbon 
monoxide and the reduction of nitrogen monoxide, using lattice gas reaction 
models. Only brief comments are given on other reactions such as the 
oxidation of hydrogen, ammonia  synthesis, etc.  
Also, a discussion of relevant experiments is presented
and measurement are compared with the numerical results. 
Furthermore, promising areas for further research and open questions 
are also addressed. 
\end{abstract}
\maketitle



	












       


\newpage

\section{Introduction}
\label{introdu}
The study and understanding of heterogeneously catalyzed
reactions is a field that contains an enormous wealth of
still unclear or even completely unexplained phenomena.
This scenario leads to an exciting and challenging domain
for investigation and fundamental research. Furthermore,
the occurrence of many complex and fascinating physical and
chemical phenomena, such as pattern formation and
self-organization \cite{A,B}, regular and irregular
kinetic oscillations \cite{A,B,ron,ger,ronert,david}, propagation and
interference of chemical waves and spatio-temporal
structures \cite{xx1,xx2}, the transition into chaotic
behaviour \cite{zz,caos}, fluctuation-induced transitions \cite{evim},
irreversible phase transitions (IPT's)
\cite{rev1,rev2,rev3,rev4,rev5}, etc, has attracted the
attention of many scientists. 

In addition to the basic interest, heterogeneous catalysis
is a field of central importance for numerous
industrial (e.g. synthesis of ammonia, sulfuric and nitric
acids, cracking and reforming processes of hydrocarbons, etc.)
and practical (e.g. catalytic control of environmental
pollution such as the emission of $CO$, $NO_{x}$, $SO_{2}$,
$O_{3}$, etc.) applications. Furthermore, information technology,
material science, corrosion, energy conversion, ecology and
environmental sciences, etc. are some fields whose rapid
growth is somehow based on the recent progress in the study 
of heterogeneous reactions occurring on surfaces and interfaces.

It should be noticed that recent developments of experimental
techniques such as Scanning Tunneling Microscopy (STM),
Low Energy Electron Diffraction (LEED), High
Resolution Electron Energy Loss Spectroscopy (HREELS), Ultraviolet
Photoelectric Spectroscopy (UPS), Photoelectron Emission Microscopy (PEEM),
etc. \cite{chr,bloc,bloc1}, just to quote few of them, allows the scientists to
gather detailed physical and chemical information about surfaces,
adsorbates and reaction products. Within this context, the
STM based measurement of the reaction rate parameters at a
microscopic level for the catalytic oxidation of $CO$ \cite{nater}
is a clear example of the progress recently achieved. 
Remarkably, the measured parameters agree very well with those
previously obtained by means of macroscopic measurements.
Also, all elementary steps of a chemical reaction
have been induced on individual molecules in a controlled
step-by-step manner with the aid of STM techniques \cite{steps}.
Furthermore, very recently, the oxidation of $CO$ on $Pt(110)$ 
was studied by means of STM techniques inside a high-pressure
flow reactor, i.e. under semirealistic conditions as compared with
those prevailing in the actual catalytic process \cite{hipr}. 
It is interesting to notice that a new reaction
mechanism, not observed when the reaction takes place under low 
pressure, has been identified \cite{hipr}.
Due to this stimulating generation of accurate experimental
information, the study of catalyzed reaction systems is
certainly a challenging scientific field for the development
and application of analytical methods, theories and numerical
simulations.

Within this context, the aim of this report is to review
recent progress in the understanding of IPT's occurring
in various lattice gas reaction systems (LGRS). 
It should be noticed that LGRS models are crude approximations
of the actual (very complex) catalytic processes. However,
from the physicist's point of view, the LGRS approach
is widely used because it is very useful to gain insight
into far-from equilibrium processes. In fact, due to the lack of a
well-established theoretical framework, unlike the case of their
equilibrium counterpart, the progress of the statistical mechanics
of systems out of equilibrium relies, up to some extent,
on the study and understanding of simple models. So, in
the case of LGRS, one renounces to a detailed description of 
the catalyzed reaction and, instead, the interest is focused 
on the irreversible critical behaviour of archetype models
inspired in the actual reaction systems.

Keeping these concepts in mind, the review will be devoted 
to survey the theoretical development
of the field during the last decade and to discuss promising 
areas for further research as well as open questions.

\subsection{Heterogeneous Catalysis and Irreversible Reaction Processes}

In most cases, heterogeneously catalyzed reactions proceed
according to well-established elementary steps. The first
one comprises trapping, sticking and adsorption. Gaseous reactant atoms
and/or molecules are trapped by the potential well of the surface. This
rather weak interaction is commonly considered as a physisorbed precursor
state. Subsequently, species are promoted to the chemisorbed state where
a much stronger interaction potential is activated. Particularly
important from the catalytic point of view is that molecules frequently
undergo dissociation, e.g. $N_{2}$, $O_{2}$, $H_{2}$, etc, which 
is a process that frees highly reactive atomic species on the surface. 
Sticking and adsorption processes
depend on the surface structure (both geometric and electronic).
In some cases, chemisorption of small atoms and molecules may
induce the reconstruction of the surface. This effect, coupled to
structure dependent sticking coefficients, may lead to the occurrence
of collective phenomena such as oscillations \cite{ron,ger,ronert}.

After adsorption, species may diffuse on the surface or, eventually,
become absorbed in the bulk. Due to collisions between adsorbed species of
different kind the actual reaction step can occur. Of course, this step
requires that energetic and spatial constraints be fulfilled.
The result of the reaction step is the formation of a product molecule. This
product can be either an intermediate of the reaction or its final output.

The final step of the whole reaction process is the desorption of
the products. This step is essential not only for the practical purpose of
collecting and storing the desired output, but also for the regeneration
of the catalytic active sites of the surface. Most reactions have at least
one rate limiting step, which frequently makes the reaction prohibitively
slow for practical purposes when, for instance, it is intended in 
an homogeneous (gas
or fluid) medium. The role of a good solid-state catalyst is to obtain
acceptable output rate of the products. Reactions occurring in this way are
commonly known as heterogeneously catalyzed.

At this stage and in order to illustrate the above-mentioned elementary steps,
it is useful to point our attention to a
specific reaction system. For this purpose, the catalytic
oxidation of carbon monoxide,
namely $2 CO + O_{2} \rightarrow 2 CO_{2}$, which is likely the
most studied reaction system, has been selected. It
is well known that such reaction proceeds according to the
Langmuir-Hinshelwood mechanism \cite{eng,eng1}, i.e with both 
reactants adsorbed on the catalyst's surface 
\begin{equation}
CO(g) + S \rightarrow CO(a)
\label{adco}
\end{equation}
\begin{equation}
O_2(g) + 2S \rightarrow 2O(a)
\label{ado}
\end{equation}
\begin{equation}
CO(a) + O(a) \rightarrow CO_{2}(g)
\label{reac}
\end{equation}
\noindent where $S$ is an empty site on the surface, while ($a$)
and ($g$) refer to the adsorbed and gas phase, respectively.
The reaction takes place with the catalyst, e.g $Pt$,
in contact with a reservoir of $CO$ and $O_{2}$
whose partial pressures are $P_{CO}$ and $P_{O_{2}}$,
respectively.

Equation (1) describes the irreversible molecular adsorption
of $CO$ on a single site of the catalyst's surface. It is known
that at under suitable temperature and pressure reaction
conditions, $CO$ molecules diffuse on the surface. Furthermore,
there is a small probability of $CO$ desorption that increases
as the temperature is raised \cite{ron}.

Equation (2) corresponds to the irreversible adsorption
of $O_{2}$ molecules that involves the dissociation of
such species and the resulting $O$ atoms occupy two
sites of the catalytic surface. Under reaction conditions
both the diffusion and the desorption of oxygen are negligible.
Due to the high stability of the $O_{2}$ molecule the
whole reaction does not occur in the homogeneous phase
due to the lack of $O_{2}$ dissociation. So, equation (2)
dramatically shows the role of the catalyst that makes
feasible the rate limiting step of the reaction.

Finally, equation (3) describes the formation of the
product ($CO_{2}$) that desorbs from the catalyst's
surface. This final step is essential for the regeneration
of the catalytic active surface.

\subsection{Absorbing States and Irreversible Phase Transitions (IPT's)}
\label{absstate}
Assuming irreversible adsorption-reaction steps, as in the
case of equations (1-3), it may be expected that on the limit
of large $P_{CO}$ and small $P_{O_{2}}$
(small $P_{CO}$ and large $P_{O_{2}}$) values, the surface
of the catalyst would become saturated by $CO$ ($O_{2}$)
species and the reaction would stop. In fact, the surface of the
catalyst fully covered by a single type of species, where further
adsorption of the other species is no longer possible, corresponds
to an inactive state of the system. This state is known as
`poisoned', in the sense that adsorbed species on the surface
of the catalyst are the poison that causes the reaction to stop.
Physicists used to call such state (or configuration)
`absorbing' because a system can be trapped
by it forever, with no possibility of escape \cite{MAM}.

These concepts are clearly illustrated in figure 1,
which shows plots of the rate of $CO_{2}$ production ($R_{CO_{2}}$)
and the surface coverage with $CO$ and $O_{2}$
($\theta_{CO}$ and $\theta_{O}$, respectively),
versus the partial pressure
of $CO$ ($P_{CO}$), as obtained using the Ziff-Gulari-Barshad
(ZGB) lattice gas reaction model \cite{zgb}. Details on the ZGB model will
be discussed extensively below, see also \cite{rev4,rev5}.
For $P_{CO} \leq P_{1CO} \cong 0.38975$ the surface
becomes irreversibly poisoned by $O$ species with
$\theta_{CO} = 0$,  $\theta_{O} = 1$ and $R_{CO_{2}} = 0$.
In contrast, for $P_{CO} \geq P_{2CO} \cong 0.5255$ the catalyst
is irreversibly poisoned by $CO$ molecules with 
$\theta_{CO} = 1$,  $\theta_{O} = 0$ and $R_{CO_{2}} = 0$.
These poisoned states are absorbing and the system cannot escape
from them. However, as shown in figure 1, between these absorbing
states there is a reaction window, namely for
$P_{1CO} < P_{CO} < P_{2CO}$, such that a steady state
with sustained production of $CO_{2}$ is observed.

\begin{figure}
\begin{center}
\epsfxsize=6cm
\epsfysize=6cm
\epsfbox{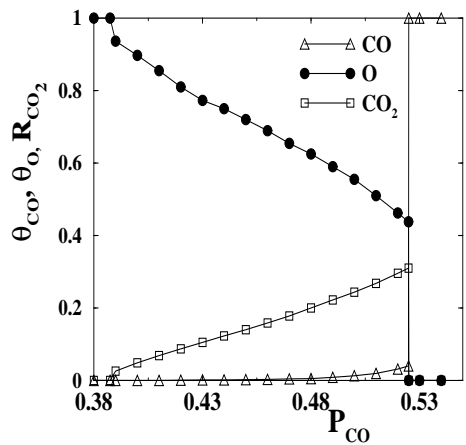}
\end{center}
\caption{Phase diagram of the ZGB model obtained using the Standard
Ensemble, showing the dependence of the surface coverage with
$CO$ ($\theta_{CO}$) and  Oxygen ($\theta_{O}$), and the rate 
of $CO_{2}$ production ($R_{CO_{2}}$) on the partial pressure
of $CO$ ($P_{CO}$) in the gas phase. Irreversible phase transitions
occurring at $P_{1CO} \simeq 0.38975$ (second-order) and 
$P_{2CO} \simeq 0.5255$ (first-order) can clearly be observed.}
\label{Figure1}
\end{figure}

It is worth mentioning that starting from the reactive regime
and approaching the oxygen absorbing state, all quantities of
interest change smoothly until they adopt the values corresponding
to the absorbing state. This behaviour typically corresponds to a
second-order irreversible phase transition (IPT). The transition
is irreversible because when the control parameter ($P_{CO}$ in
this example) is tuned into the absorbing state the system
becomes trapped by it forever. This behaviour is in contrast
to that observed for second-order reversible phase transitions,
such as the order-disorder transition of the Ising
ferromagnet in the absence of an external magnetic field,
where it is possible to change reversibly from one phase to
the other, simply tuning the control parameter \cite{kur1,kur2}.
For second-order IPT's, as in the case of their reversible
counterparts, it is possible to define an order parameter, which
for the former is 
given by the concentration of minority species ($\theta_{CO}$,
in the case of the second-order IPT of the catalytic oxidation
of $CO$). Furthermore, it is known
that $\theta_{CO}$ vanishes according to a power law upon
approaching the critical point \cite{meakin}, so that

\begin{equation}
\theta_{CO} \propto (P_{CO} - P_{1CO})^{\beta} ,
\end{equation}
\noindent where $\beta$ is the order parameter critical exponent
and $P_{1CO}$ is the critical point.

Remarkably, the behaviour of the system is quite different upon
approaching the $CO$ absorbing state from the reactive regime
(see figure 1). In this case, all quantities of interest exhibit a
marked discontinuity close to $P_{2CO} \cong 0.5255$. This is
a typical first-order IPT and $P_{2CO}$ is the coexistence point.

Experimental results for the catalytic oxidation of carbon monoxide
on $Pt(210)$ and $Pt(111)$ \cite{bloc,bloc1} are in qualitative agreement with
simulation results of the ZGB model, it follows from the
comparison of figures 1 and 2. A remarkable agreement is the (almost)
linear increase in the reaction rate observed when the $CO$ pressure
is raised and the abrupt drop of the reactivity when a certain
`critical' pressure is reached. In spite of the similarities observed, 
two essential differences are
worth discussing: i) the oxygen-poisoned phase exhibited by
the ZGB model within the $CO$ low-pressure regime is not observed
experimentally. Therefore, one lacks experimental evidence
of a second-order IPT. ii) The $CO-$rich phase exhibiting low reactivity
found experimentally resembles the $CO-$poisoned state predicted
by the ZGB model. However, in the experiments the nonvanishing
$CO-$desorption probability prevents the system from entering  into a truly
absorbing state and the abrupt, `first-order like' transition,
shown in figure 2 is actually reversible. Of course, these and other
disagreements are not surprising since the lattice gas reaction model,
with a single parameter, is a simplified approach to the
actual catalytic reaction that is far more complex.

\begin{figure}
\vskip 1.5 true cm
\begin{center}
\centerline{
\epsfxsize=8cm
\epsfysize=6cm
\epsfbox{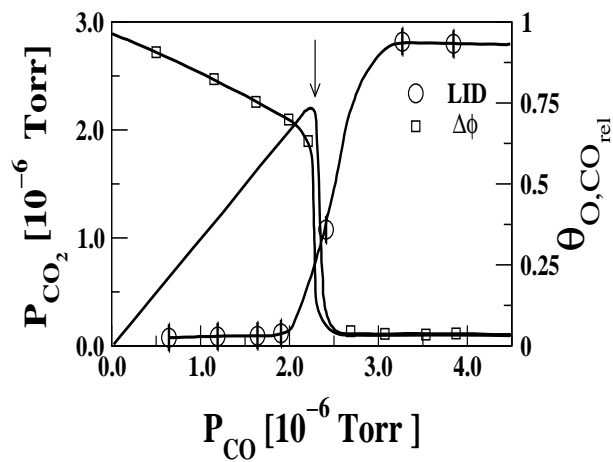}}
\end{center}
\caption{Experimental data corresponding to the catalytic
oxidation of carbon monoxide on Pt(210) obtained at $T = 500 K$,
keeping the oxygen pressure constant at  
$P_{O} = 2.0 \times 10^{-6} Torr$ and tuning the $CO$ pressure
(horizontal axis). The left vertical axis shows the partial pressure
of $CO_{2}$ that is proportional to the rate of 
$CO_{2}$ production ($R_{CO_{2}}$). The right vertical axis shows  
the surface coverage with $CO$ ($\theta_{CO}$) and 
oxygen ($\theta_{O}$), measured relative to their respective maxima.
$CO$ coverages were measured by means of the absorbate
induced work function changes $\Delta \Phi$, while oxygen coverages were
determined using Laser-Induced Desorption (LID). The transition point is 
shown with an arrow. Adapted from references \cite{bloc,bloc1}.}
\label{Figure2}
\end{figure} 

At this stage it is worth warning the reader that
due to the non-Hamiltonian nature of the ZGB model,
as well as all the lattice gas reaction systems that
will be treated hereafter, one lacks a thermodynamic
quantity, such as the free energy, in order to draw a
sharper definition of the order of the transitions.
Even more, in contrast to their reversible counterpart,
the field of IPT's lacks a well-established theoretical
framework.
Also, it is worth mentioning that systems with absorbing
states are clearly out of equilibrium. In fact,
the transition rate out of the absorbing state is
zero, such as those configurations cannot fulfill
standard detailed balance requirements. Therefore,
the study of the critical behaviour of these systems
must involve IPT's between reactive and absorbing phases.
Due to these circumstances, the study of IPT's
represents a theoretical challenge within the more general
goal of modern physics given by the development of
the statistical mechanics of nonequilibrium systems.

It should be recognized that the study of absorbing states
and IPT's is almost ignored in the courses of statistical mechanics,
in spite of the fact that they abound in physics,
chemistry, biology and other disciplines including
sociology. Some typical examples include the spreading
of epidemics through a population, the propagation
of rumors in a society \cite{damian}, the spreading of fire through
a forest \cite{clar,evaff}, coexistence with extinction transitions in
prey-predator systems \cite{ale} and, of course, catalytic and autocatalytic
chemical reactions. From a more general point of view,
absorbing states are expected to occur in situations
where some quantity of interest can proliferate or
die out ( e.g the fire in the forest), without any
possibility of spontaneous generation (e.g due to
the rays of an electrical storm). The underlying
physics involves the competition between proliferation
and death of a certain quantity. Proliferation is
associated with the active (reactive) phase of the system,
while inactivity (poisoning) characterizes the
absorbing phase.

\section{ Theoretical Approaches}

There are currently three basic approaches for the theoretical
modeling of surface reactions: i) {\it ab-initio} calculations,
ii) analytic approaches and iii) stochastic models. 
The {\it ab-initio} method is usually
implemented via density functional theory approaches and due
to the huge computational requirements, the calculations are
restricted to few atomic layers of the catalysts and a very
reduced catalyst's surface (of the order of $1 nm^{2}$).
This approach frequently allows the study of a single adsorbed
species or a reactive pair only. Consequently, the
study of macroscopic surface systems incorporating statistical
effects, as requested for the case of critical phenomena,
is impossible at present. So, this approach will not be
further discussed here.

On the other hand, stochastic models can account for
fluctuations in large systems. So, they are used
to deal with a large number of collective phenomena occurring
in reaction systems that are not only restricted to IPT's,
but also involve spatio-temporal structures \cite{dani,nl}, 
chemical waves \cite{jimn,dios,evawaves}, kinetic 
oscillations \cite{vigi,iner,aren,evla,evpre,evjcp,kuzo}, 
the transition to chaos \cite{zz,caos}, etc..

Broad experience gained in the treatment of equilibrium
systems has shown that Monte Carlo simulations \cite{kur1,kur2} and
Renormalization Group (RG) analysis of classical
field-theoretical models \cite{cardy} are among the most useful tools
for the treatment of phase transitions and critical
phenomena. A much more reduced experience, obtained during
the last decade, indicates that, after some adaptation,
similar techniques can also been employed to deal with IPT's.

\subsection{The Monte Carlo Method}

Monte Carlo (MC) simulations of heterogeneously catalyzed reactions
can be considered the computational implementation of microscopic
reaction mechanisms. In fact, such mechanisms are the `rules'
of the computer algorithm. Of course, the
operation of the rules may lead to the development of correlations, while
stochastic fluctuations are inherent to the method.

For the practical implementation of the MC method, the catalyst's surface
is replaced by a lattice. Therefore, lattice gas reaction models
are actually considered.  For this reason, the method
often faces the limitations imposed by the size of the lattices used.
In some particular cases, e.g. when studying
second-order phase transitions, this shortcoming can be overcome appealing
to the well-established finite-size-scaling theory \cite{evaprb,gri}. 
Also, very
often one can develop extrapolation methods that give reliable results for
the thermodynamic limit, i.e. infinite lattices. Another limitation arises
when the diffusion rate of the adsorbed species is very large. In this case,
most of the computational time has to be devoted to the diffusion process
while the quantity of interest, namely the number of reaction events,
becomes negligible. This drawback may be overcome implementing a mixed
treatment: mean-field description of the diffusion and MC simulation of the
reaction \cite{sil,jimi2}. This approach may become an interesting and powerful
tool in the near future.

MC simulations of dynamic and kinetic
processes are often hindered by the fact that the Monte Carlo time, usually
measured in Monte Carlo time steps, is somewhat proportional to the actual
time. So, direct comparison with experiments becomes difficult. However,
very recently a more sophisticated implementation of the MC method has been
envisioned: namely the Dynamic Monte Carlo (DMC) approach that 
incorporates the actual time dependence of the processes allowing direct
comparison with experiments \cite{nl,kris,bme,apj}. Further developments and
applications of the DMC method are a promising field of research.

Within this context, the following subsections describe 
different MC approaches suitable for the study of
IPT's in reaction systems, namely the Standard Ensemble, the Constant
Coverage Ensemble and the Epidemic Method. Furthermore, the standard
finite-size scaling theory adapted to the case of second-order IPT's
and they application to MC data are also discussed. 

\subsubsection{The Standard Ensemble}
\label{estensam}
In this ensemble the catalyst is assumed to be in contact with an
infinitely large reservoir containing the reactants in the gas phase.
Adsorption events are treated stochastically neglecting energetic
interactions. The reaction between reactants takes place on
the surface of the catalyst, i.e. the so-called Langmuir-Hinshelwood
mechanism. After reaction, the product is removed from the surface
and its partial pressure in the gas phase is neglected, so that
readsorption of the product is not considered.

In order to further illustrate the practical implementation of the
Standard Ensemble let us describe the simulation method of the
lattice gas reaction version of the catalytic oxidation of $CO$
(equations (1-3)) according to the ZGB model \cite{zgb} on the two-dimensional
square lattice. The Monte Carlo algorithm is as follows:

i) $CO$ or $O_{2}$ molecules are selected at random with relative
probabilities $P_{CO}$ and $P_{O}$, respectively. These
probabilities are the relative impingement rates of both
species, which are proportional to their partial pressures
in the gas phase in contact with the catalyst. Due to the
normalization, $P_{CO} + P_{O} = 1$, the model has a single
parameter, i.e. $P_{CO}$. If the selected species
is $CO$, one surface site is selected at random, and if that site
is vacant, $CO$ is adsorbed on it according to equation (1).
Otherwise, if that site is occupied, the trial ends and a
new molecule is selected. If the selected species is $O_{2}$,
a pair of nearest-neighbor sites is selected at random and the
molecule is adsorbed on them only if they are both vacant, as
requested by equation (2).

ii) After each adsorption event, the nearest-neighbors of the
added molecule are examined in order to account for the reaction
given by equation (3). If more than one [$CO(a),O(a)$] pair
is identified, a single one is selected at random and
removed from the surface.

The phase diagram of the ZGB model, as obtained using the
Standard Ensemble, is shown in figure 1 and will be further
discussed in \Sref{zgbsection}.

\subsubsection{The Constant Coverage Ensemble}

Monte Carlo  simulations using the Constant Coverage (CC) ensemble,
as early proposed by Ziff and Brosilow \cite{zb}, are likely the
most powerful method available for the study of first-order IPT's.

In order to implement the CC method, first a stationary
configuration of the system has to be achieved using the
Standard Ensemble algorithm, as described in \Sref{estensam}
For this purpose, one selects a value of the parameter close to
the coexistence point but within the reactive regime. 
After achieving the stationary state,
the simulation is actually switched to the CC method. Then,
one has to maintain the coverage with the majority species
of the corresponding absorptive state as constant as possible
around a prefixed value. This goal is achieved regulating
the number of adsorption events of competing species.
The ratio between the number of attempts made
trying to adsorb a given species and the total number of
attempts is the measure of the corresponding effective
partial pressure. In this way, in the CC ensemble the
coverage assumes the role of the control parameter.

In the case of the ZGB model, the stationary state is usually
achieved for $P_{CO} = 0.51$. So, in order to maintain
the $CO$ coverage close to the desired value $\theta_{CO}^{CC}$,
oxygen ($CO$) adsorption attempts take place whenever
$\theta_{CO} > \theta_{CO}^{CC}$ ($\theta_{CO} < \theta_{CO}^{CC}$).
Let ${\cal N}_{CO}$ and ${\cal N}_{OX}$ be the number of
carbon monoxide and oxygen attempts, respectively.
Then, the value of the `pressure' of $CO$  in the
$CC$ ensemble ($P_{CO}^{CC}$) is determined just as the ratio 
$P_{CO}^{CC} = \frac{{\cal N}_{CO}}{{\cal N}_{CO} + {\cal N}_{OX}}$.
Subsequently, the coverage is increased by small amount,
say $\Delta\theta_{CO}^{CC}$. A transient period $\tau_{P}$ is
then disregarded for the proper relaxation of the system to the  
new $\theta_{CO}^{CC}$ value, and finally averages of $P_{CO}^{CC}$
are taken over a certain measurement time $\tau_{M}$.

In the original CC algorithm of Brosilow and Ziff \cite{zb}
the coverage of $CO$ was increased stepwise up to a certain
maximum value ($\theta_{CO}^{max}$) and the set of points
($\theta_{CO}^{CC}, P_{CO}^{CC}$) were recorded.
However, later on it has been suggested \cite{weJPA}
that it would be convenient to continue the simulations after
reaching $\theta_{CO}^{max}$ by {\bf decreasing} 
$\theta_{CO}^{CC}$ stepwise until $P_{CO}^{CC}$
reaches a value close to the starting point, namely
$P_{CO}^{CC}=0.51$. In fact, this procedure would allow
to investigate one possible hysteretic effects at first-order
IPT's, which are expected to be relevant as follows from
the experience gained studying their counterpart in equilibrium
(reversible) conditions.

\subsubsection{Finite-Size Scaling Approach to Second-Order IPT's}

Since true phase transitions only occur on the thermodynamic
limit and computer simulations are always restricted to finite samples,
numerical data are influenced by rounding and shifting effects
around pseudo critical points \cite{kur1,kur2}. Within this context
the finite-size scaling theory \cite{cardy,barber} has become a powerful
tool for the analysis of numerical results allowing the determination
of critical points and the evaluation of critical exponents.
All this experience  gained in the study of reversible
critical phenomena under equilibrium conditions can be
employed in the field of second-order IPT's. 

In order to perform a finite-size scaling analysis close to a
second-order ITP \cite{evaprb,gri} in a reaction system it 
is convenient to take
the concentration of the minority species on the surface,
($\theta_{M}$), as an order parameter. By analogy to
reversible second-order transitions on assumes that

\begin{equation}
\theta_{M} \propto (P - P_{c})^{\beta} ,
\label{orpa}
\end{equation}

\noindent where $\beta$ is the order parameter critical exponent and
$P_{c}$ is the critical value of the control parameter $P$. Approaching
$P_{c}$ from the reactive regime, the characteristic length scale of
the system given by the correlation length $\xi_{\bot}$
diverges according to 

\begin{equation}
\xi_{\bot} \propto (P - P_{c})^{-\nu_{\bot}} ,
\label{corlen}
\end{equation}

\noindent where $\nu_{\bot}$ is the correlation length exponent
in the space direction.

For finite samples and close to the critical region, the
concentration of minority species will depend on two
competing lengths, namely $\theta_{M}(L, \xi_{\bot})$,
and the scaling hypothesis assumes

\begin{equation}
\theta_{M}(L, \xi_{\bot}) =
L^{-\beta/{\nu_{\bot}}} F[(P - P_{c}) L^{1/\nu_{\bot}}] ,
\end{equation}

\noindent where equation (\ref{corlen}) has been used and $F$ is a suitable
scaling function. Just at $P_{c}$, one has 

\begin{equation}
\theta_{M}(L, P_{c})  \propto L^{-\beta/{\nu_{\bot}}}  ,
\end{equation}

\noindent and

\begin{equation}
F(x)  \propto x^{\beta} ,\qquad (x \rightarrow \infty),    
\end{equation}

\noindent such that equation (\ref{orpa}) is recovered when
$L \rightarrow \infty$ in the critical region.

As anticipated by the adopted notation, second-order IPT's
exhibit spatio-temporal anisotropy, so that the
correlation length in the time direction is given by

\begin{equation}
\xi_{\|} \propto (P - P_{c})^{-\nu_{\|}} ,
\end{equation}

\noindent where $\nu_{\|}$ is the corresponding correlation
length exponent.

Examples of the application of finite-size scaling to
various reaction systems can be found in the literature
\cite{evaprb,gri,baw,baw1}. However, a more accurate method for the
evaluation of critical exponents is to combine finite-size 
scaling of stationary quantities and dynamic scaling,
as will be discussed just below.

\subsubsection{The Epidemic Method and Dynamic Scaling}
\label{epidemias}
In order to obtain accurate values of the critical point and
the critical exponents using the Standard Ensemble and applying
finite-size scaling, it is necessary to perform MC simulations
very close to the critical point. However, at criticality and
close to it, due to the large fluctuations always present in
second-order phase transitions, any finite system
will ultimately become irreversibly trapped by the
absorbing state. So, the measurements are actually 
performed within metastable states facing two competing 
constraints: on the one hand the measurement time has to be long enough
in order to allow the system to develop the corresponding
correlations and, on the other hand, such time must be short enough
to prevent poisoning of the sample. This shortcoming can be
somewhat healed by taking averages over many samples
and disregarding those that have been trapped by the
absorbing state. However, it is difficult to avoid
those samples that are just evolving to such absorbing state,
unless that fluctuations are suppressed by comparing two
different samples evolving through the phase space following
very close trajectories \cite{DS}. In view of these
shortcomings, experience indicates that the best
approach to second-order IPT's is to complement finite-size 
scaling of stationary quantities, as obtained with the
Standard Ensemble, with epidemic simulations.

The application of the Epidemic Method (EM) to the
study of IPT's has become a useful tool for the evaluation 
of critical points, dynamic critical exponents and eventually for the 
identification of universality classes \cite{Jens4,Grass1,Grass2}. 
The idea behind the EM is to initialize the
simulation using a configuration very close to the 
absorbing state. Such a configuration can be achieved
generating the absorbing state using the Standard Ensemble and,
subsequently, removing some species from the center
of the sample where a small patch of empty sites is
left. In the case of the ZGB model this can be done by filling the
whole lattice with $CO$, except for a small patch.
Patches consisting of 3-6 neighboring empty sites
are frequently employed, but it is known that the asymptotic results
are independent of the size of the initial patch.
Such patch is the kernel of the subsequent epidemic.

After the generation of the starting configuration,
the time evolution of the system is followed using the
Standard Ensemble as already described in \Sref{estensam}.
During this dynamic process the following quantities are recorded:
(i) the average number of empty sites ($N(t)$),
(ii) the survival probability $P(t)$, which is the probability
that the epidemic is still alive at time $t$, and
(iii) the average mean square distance, $R^{2}(t)$,
over which the empty sites have spread. Of course,
each single epidemic stops if the sample is trapped in the
poisoned state with $N(t) = 0$ and, since these events may happen
after very short times (depending on the patch size), results
have to be averaged over many different epidemics. It should be 
noticed that $N(t)$ ($R^2(t)$) is averaged over all (surviving)
epidemics. 

If the epidemic is performed just at critically, a power-law
behaviour (scaling invariance) can be assumed and the
following ans\"atze are expected to hold,
 
\begin{equation}
N(t)\,\propto\,t^{\eta}  ,
\label{nu}
\end{equation}

\begin{equation}
P(t)\,\propto\,t^{-\delta} ,
\label{so}
\end{equation}

\noindent and

\begin{equation}
R^{2}(t)\,\propto\,t^{z} ,
\label{R2}
\end{equation}

\noindent where $\eta$, $\delta$ and $z$ are dynamic critical exponents.
Thus, at the critical point log-log plots of $N(t)$, $P(t)$ and $R^2(t)$
will asymptotically show a straight line behaviour, while off-critical
points will exhibit curvature. This behaviour allows the
determination of the critical point and from the slopes of the plots
the critical exponents can also be evaluated quite accurately \cite{Grass1}.

Using scaling arguments, it has been shown that the following
relationship \cite{Grass1}

\begin{equation}
(d + 1)z = 4 \delta + 2 \eta ,
\label{anzatGdT}
\end{equation}

\noindent holds, allowing the evaluation of one exponent
as a function of the other two.

The validity of equations (\ref{nu}), (\ref{so}) and (\ref{R2}) 
for second-order IPT's is
very well established. Furthermore, the observation 
of a power-law behaviour for
second-order IPT's is in agreement with the 
ideas developed in the study of equilibrium (reversible)
phase transitions: scale invariance reflects the existence
of a diverging correlation length at criticality. 
It should be mentioned that the EM can also be applied
to first-order IPT's close to coexistence. However,
since it is well known that in the case of first-order reversible
transitions correlations decay exponentially, preventing
the occurrence of scale invariance,
equations (\ref{nu}), (\ref{so}) and (\ref{R2}) have to be
modified. Recently, the following anzats has been proposed \cite{weJPA}

\begin{equation}
N(t)\propto(\frac{t}{\tau})^{-\eta^{eff}} \exp[-(\frac{t}{\tau})]
\label{anz}
\end{equation}

\noindent where $\eta^{eff}$ is an effective exponent and $\tau$ sets
a characteristic time scale. So, equation (\ref{anz}) combines
a power-law behaviour for $t\rightarrow\,0$ with an exponential
(asymptotic) decay.

It should also be mentioned that the whole issue of the occurrence of
power-law behaviour at equilibrium first-order transitions is a bit
more complex than the simplified arguments used above. For example,
scaling behaviour inside the coexistence zone has been observed for the
liquid-gas phase transitions using finite systems. However, this
scaling disappears on the thermodynamic limit \cite{gul}. Also, when a 
first-order line ends in a second-order critical point, the system
frequently displays several decades of critical behaviour (before the
exponential roll-off) even when measurements are performed quite a
bit beyond the critical point \cite{bark}.
 
\subsection{Analytical Methods}

In the field of reversible critical phenomena, the most effective
analytical tool for the identification of universality classes is the
Renormalization Group  analysis of classical field-theoretical
models using coarse-grained Ginzburg-Landau-Wilson free-energy
functionals \cite{GLW}. While reaction systems exhibiting IPT's do not have
energy functionals, they can often be treated on the coarse-grained
level by means of phenomenological Langevin equations. Stochastic partial
differential equations of this kind are the
natural formalism to analyze critical properties of irreversible
systems with absorbing states, as will be discussed below.

On the other hand, Mean-Field modeling using ordinary
differential equations (ODE) is a widely used method \cite{rev3}
for the study of first-order IPT's (see also below).
Further extensions of the ODE framework, to include
diffusional terms are very useful and, have
allowed the description of spatio-temporal patterns
in diffusion-reaction systems 
\cite{dani}. However, these methods are essentially limited for the
study of second-order IPT's because they always consider average
environments of reactants and adsorption sites, ignoring stochastic
fluctuations and correlations that naturally emerge in
actual systems.

\subsubsection{The Langevin Formulation}
\label{lang}
The Langevin equation for a single-component field $\phi({\bf x}, t)$
can be written as \cite{MAM}

\begin{equation}
\frac{\partial \phi({\bf x}, t)}{\partial t} = 
F_{x}(\{ \phi \}) + G_{x}(\{ \phi \})\eta({\bf x},t) ,
\label{lan}
\end{equation}

\noindent where $F_{x}$ and $G_{x}$ are functionals of $\phi$,
and $\eta$ is a Gaussian random variable with zero mean, such
that the only nonvanishing correlations are
$\langle \eta({\bf x},t) \eta({\bf x'},t')\rangle =
D \delta({\bf x} - {\bf x'}) \delta(t - t')$.
Equation (\ref{lan}) is purely first-order in time since
it is aimed to describe critical properties where
small-wavenumber phenomena dominate the physical behaviour. 

The goal of the coarse-grained Lagevin representation is to capture
the essential physics of microscopic models on large
length and time scales. Therefore, $F_{x}$ and $G_{x}$
are taken to be analytical functions of $\phi({\bf x}, t)$
and its spatial derivatives. In this way, coarse graining
smooths out any nonanalyticity of the underlying
microscopic dynamic model.

At this stage, it is clear that one has an infinite number of
possible analytical terms in $\phi$ and its space derivatives
that can be placed on the right hand side term of equation (\ref{lan}).
Following the ideas of Landau and Lifshitz, it is expected that
symmetry considerations would suffice to determine the
relevant terms of $F_{x}$ and $G_{x}$. In fact, these functional
must include all analytic terms consistent with the
symmetry of the microscopic model and no term with
lower symmetry can be included. The coefficients of
these terms are simply taken as unknown parameters or
constants to be determined phenomenologically. This
assumption is consistent with the fact that the typical
IPT's, wich are intended to be described with
equation (\ref{lan}), can be characterized by a set of
critical exponents defining their universality class, which
do not depend on the values of the parameters.

Depending on the behaviour of the noise amplitude
functional $G_{x}$, two types of physical systems
can already be treated:

a) Systems without absorbing states. In this case
$G_{x}(\phi)$ approaches a constant value. Using
Renormalization Group arguments it can be shown that
higher powers of $\phi$ and its derivatives in $G_{x}$
are irrelevant for the description of the critical
behaviour and therefore, they can be neglected.
In this way, the Langevin equation has a simple additive
Gaussian noise of constant amplitude.
Assuming that $F_{x}(\phi)$ can be written as

\begin{equation}
F_{x}(\phi) = \Gamma( \nabla^{2} \phi - r \phi + u \phi^{3}) ,
\end{equation}

\noindent where $r$ and $u$ are constants, the best known
example results in the Ginzburg-Landau-Wilson functional derivative
of free energy functional from the Ising model, e.g.

\begin{equation}
{\cal{H} (\phi)}= \int d{\bf x} [ (\nabla \phi)^{2} + r \phi^{2}/2 + u \phi^{4})/4 ],
\end{equation}

\noindent such that the Langevin equation becomes the celebrated
time dependent Ginzburg-Landau equation. This equation
is very useful for the treatment of equilibrium critical
phenomena and has been quoted here for the sake of completeness only.
In fact, in this article, our interest is focused on far-from
equilibrium systems, for which $F_{x}(\phi)$ cannot be expressed
as a functional derivative. Within this context the best known
example is the Kardar-Parisi-Zhang (KPZ) equation introduced
for the description of the dynamic behaviour of interfaces without
overhangs. For systems with interfaces, $\phi({\bf x},t)$
is taken as the height at time $t$ of a site of the interface
at position ${\bf x}$. It has been shown that the functional

\begin{equation}
F_{x}(\phi) = c \nabla^{2} \phi  + \lambda (\nabla \phi)^{2}) ,
\end{equation}

\noindent captures the large scale physics of
moving interfaces leading to the KPZ equation \cite{kpz}

\begin{equation}
\frac{\partial \phi}{\partial t}  = c \nabla^{2}\phi + 
\lambda (\nabla \phi)^{2}) + \eta({\bf x},t).
\label{kkppzz}
\end{equation}

Recent studies have demonstrated that the KPZ equation
holds for the treatment of a wide variety of physical
situations \cite{bara} including the description of
the interfacial behaviour of reactants close to
coexistence in systems exhibiting first-order IPT's \cite{jimn,dios},
for further details see also \Sref{zgbsection}.

b) Systems with absorbing states. In the Langevin equations
capable of describing these systems, the noise amplitude functional
$G_{x}(\phi)$ must vanish with $\phi$  so that the
state  $\phi({\bf x}) = 0$ is an absorbing state
for any functional $F_{x}$ that has no constant piece and
therefore it also vanishes with $\phi$.
It can be demonstrated that $G_{x}(\phi) \propto \phi^{\alpha}$,
for small enough values of $\phi$. Also, $\alpha$ can assume
only positive integer and half-odd-integer values.
The cases $\alpha = 1/2$ and $\alpha = 1$ are
those of interest in practice, the former includes the
directed percolation (DP) universality class (which is of
primary interest in the context of this article because it
is suitable for the
description of second-order IPT's \cite{dana}), while the
latter includes the problem of multiplicative noise.

Considering the case of DP \cite{dp1,dp2}, one has to keep in mind that
the state $\phi = 0$ must be absorbing, so that 
both functionals  $F_{x}(\phi)$  and $G_{x}(\phi)$
must vanish as $\phi \rightarrow 0$. This constraint implies
that terms independent of $\phi$ must cannot be considered.
Now, imposing the condition of symmetry inversion on the
underlying lattice (${\bf x} \rightarrow -{\bf x}$), so that
terms containing $\nabla \phi$ are forbidden, the Langevin
equation with the lowest allowed terms in $\phi$ and its
derivatives becomes

\begin{equation}
\frac{\partial \phi}{\partial t}  = c \nabla^{2}\phi - r \phi 
-u \phi^{2} + \phi^{\alpha} \eta({\bf x},t).
\label{lan1}
\end{equation}

\noindent where $c$, $r$ and $u$ are constants.

Using Renormalization Group  methods it can be shown that 
the critical dimension
for equation (\ref{lan1}) is $d_{c} = 4$ and that the critical
exponents in terms of the first order $\epsilon-$expansion
(here $\epsilon \equiv 4 - d$) are \cite{MAM}: dynamic exponent
$z^{*} = 2 - \epsilon/12$, order parameter critical exponent
$\beta = 1 - \epsilon/6$ and correlation length exponent
$\nu_{\bot} = (1 + \epsilon/8)/2$. Also, the scaling relation

\begin{equation}
\beta = \nu_{\bot} (d - 2 + \eta^{*}) ,
\label{scarel}
\end{equation} 

\noindent holds, where $\eta^{*}$ is the critical exponent of the
two-point correlation function. Therefore, only two of the
exponents are truly independent. Notice that equation (\ref{scarel})
is a nonequilibrium counterpart of the well-known
scaling relation  $2 \beta = \nu (d - 2 + \eta)$
valid for a system without absorbing states.

Apart from the scaling relations given by equation (\ref{anzatGdT})
and (\ref{scarel}), which hold for dynamic and
static exponents, the following relationships between
exponents have been found to hold \cite{MAM}:
$2 \eta^{*} = 4 - d - z^{*} \eta$,  $z^{*} = 2/z$ and
$\delta = \nu_{\|} \beta$.

It should also be noticed that IPT's subject to extra symmetries,
and thus out of the DP universality class, have been identified in 
recent years. Among then one has systems with symmetric absorbing 
states \cite{paloma}, models of epidemics with perfect 
immunization \cite{epwi,Jan}, and systems with an 
infinite number of absorbing states \cite{dd1,je}. Additional discussions
on these topics are beyond the aim of this article since these
situations do not appear in the reaction models discussed here.
For further details the reader is addressed to the recent review
of Hinrichsen \cite{haye}.

\subsubsection{Mean-Field Treatments}

The Mean Field (MF) approach to irreversible reactions neglects
spatial fluctuations and the effect of noise. So, the actual
concentrations of the reactants are replaced by
averaged values over the sites of the lattice. Differential
equations for the time dependence of such averages are
derived for specific reaction systems with different
degrees of sophistication, for instance involving one or more
site correlations.  MF equations are useful in order to
get insight into the behaviour of reaction systems close to
first-order IPT's where spatial fluctuations are expected
to be irrelevant \cite{rev3}.

In spite of the restricted application  of MF equations to
the study of second-order IPT's, it is instructive to
discuss the MF predictions of the
Langevin equation (\ref{lan1}).  In fact, replacing the field
$\phi({\bf x},t)$  by the spatial constant
$\phi(t)$ and neglecting the noise and the Laplacian terms,
equation (\ref{lan1}) becomes \cite{MAM}

\begin{equation}
d\phi/dt = -r \phi - u \phi^{2} .
\label{lanMF}
\end{equation}

Equation (\ref{lanMF}) has two long-time stationary solutions
given by the absorbing state ($\phi = 0$) and the
active regime ($\phi = -r/u$), wich 
are stable for $r > 0$ and $r < 0$.
Here $r$ plays the role of the control parameter with
a critical point at $r_{c} = 0$. The order parameter
critical exponent that governs the decay of the average density
$\langle \phi \rangle \approx |r - r_{c}|^{\beta}$ for $r \rightarrow r_{c}$
from the active phase is $\beta = 1$.

\section{Critical Behaviour of Specific Reaction Systems}

The first part of this section will be devoted to describe 
lattice gas reaction models inspired in actual catalytic reactions,
such as the catalytic oxidation of carbon monoxide (\Sref{zgbsection})
and the reaction between nitric oxide and carbon monoxide 
(\Sref{nomasco}). These models
exhibit many features characteristic of first-order IPT's that
have also been found in numerous experiments, 
such as hysteretic effects \cite{chr,hister}
and abrupt changes of relevant properties when a control parameter
is tuned around a coexistence point \cite{chr,bloc,bloc1} (see
also figure 2), etc. On view of these facts, special 
attention will be drawn to the discussion of first-order ITP's.

On the other hand, \Sref{bosta} will be mostly devoted to the discussion 
of generic models, which are not intended to describe specific reaction 
systems. Most of these models exhibit second-order ITP's.

\subsection{ The Catalytic Oxidation of Carbon Monoxide}
\label{zgbsection}
As already discussed in the introductory sections,
the catalytic oxidation of carbon monoxide is one
of the most studied reaction due to its practical
importance and theoretical interest. The simplest
approach for this reaction is the ZGB lattice gas
model \cite{zgb} as described in \Sref{introdu} and \Sref{estensam}
(for additional details see \cite{rev1,rev4,rev5}).

The phase diagram of the ZGB model (figure 1) exhibits a
second-order IPT close to $P_{1CO} \simeq 0.3874$ that
belongs to the DP universality class \cite{dana}. The dynamic
critical exponents, as evaluated using epidemic
simulations (\Sref{epidemias}), equations (\ref{nu}), 
(\ref{so}) and (\ref{R2})), are 
$\eta = 0.224 \pm 0.010$,  $\delta = 0.452 \pm 0.008$
and $z = 1.13 \pm 0.01$ (in two dimensions) \cite{Jens4,bobZ},
which, in fact, are in excellent agreement with the acepted exponents 
of the universality class of the directed percolation, namely
$\eta = 0.2295 \pm 0.0010$, $\delta = 0.44505 \pm 0.0010$ and
$z = 1.1325 \pm 0.0010$ \cite{Grass1,Grass2,muve}. The order 
parameter critical exponent,
as evaluated using the damage spreading technique \cite{DS}, is
$\beta = 0.57 \pm 0.01$, also in excellent agreement with the
DP value, namely $\beta = 0.583 \pm 0.004$ \cite{muve}.

More interesting, close to $P_{2CO} = 0.5258$,
the ZGB model exhibits a first-order IPT (see figure 1)
in qualitative agreement with experiments performed using
single crystal surfaces as catalysts (see figure 2).
As mentioned above, the nonvanishing
$CO$ desorption rate experimentally observed \cite{ron,ger}
prevents the actual catalyst system from entering in to a truly
absorbing state and the abrupt transition
shown in figure 2 is actually reversible. Further experimental
evidence on the existence of this first-order-like behaviour arises
from dynamic measurements exhibiting clear hysteretic effects,
as shown in figure 3 for the case of the $CO + O_{2}$ reaction
on a $Pt(111)$ single crystal surface \cite{hister}.
\begin{figure}
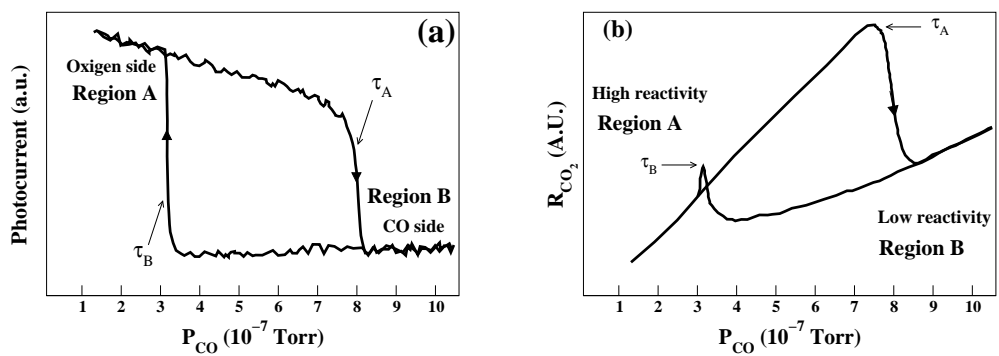

\begin{minipage}{7.0cm}
\begin{center}
\includegraphics[width=6.0cm]{Fig3a.eps}
\end{center}
\end{minipage}
\begin{minipage}{7.0cm}
\begin{center}
\includegraphics[width=6.0cm]{Fig3b.eps}
\end{center}
\end{minipage}
\caption{Experimental data corresponding to the catalytic
oxidation of carbon monoxide on Pt(111) obtained at $T = 413.2 K$,
keeping the oxygen pressure constant at  
$P_{O} = 4.0 \times 10^{-5} Torr$, while the $CO$ partial pressure is 
varied cyclically (horizontal axis). 
(a) Hysteresis in the reactant coverage
as measured by the Photoelectron Emission 
Microscopy (PEEM) \cite{la10} and (b) in the $CO_{2}$ reaction
rate. More details in the text. Adapted from reference \cite{hister}.}
\label{Figure3}
\end{figure}
Figure 3(a) shows hysteresis in the reactant coverage upon cyclic
variation of the $CO$ partial pressure (horizontal axis). 
The vertical axis shows the photocurrent measured in a 
Photoelectron Emission Microscopy (PEEM) \cite{la10} experiment.
Notice that a low (negative) photocurrent indicates an 
oxygen-rich phase (left-hand side of figure 3(a)), while 
a large photocurrent corresponds to a $CO-$rich phase 
(right-hand side of figure 3(a)). Also,
figure 3(b) shows the hysteresis in the rate of 
$CO_{2}$ production (measured using a mass spectrometer)
as a function of the $CO$ partial pressure.
When the system is in the low $CO$ pressure regime, the surface
is mostly covered by oxygen that corresponds to the 
`oxygen side' or monostable region A. Upon increasing $P_{CO}$
the reaction rate also rises until, close to $P_{CO} = \tau_{A}$,
the surface becomes covered with adsorbed $CO$. This `$CO$ side'
or monostable state B, corresponds to a low reactivity regime.
Decreasing $P_{CO}$, the system remains in this monostable
region B until, close to $P_{CO} = \tau_{B}$, it suddenly
undergoes a steep transition and rapidly returns to the initial
state of high reactivity. 
Summing up, during the hysteresis loop the system may be in
two monostable regions A ($P_{CO} < \tau_{B}$) and B  
($P_{CO} > \tau_{A}$), separated from each other by a bistable
region ($\tau_{A} > P_{CO} > \tau_{B}$) \cite{hister}.  

In view of this stimulating experimental evidence let us 
review some numerical studies on hysteretic effects
close to coexistence. 
Figure 4 shows a plot of ${\theta}_{CO}$ versus $P_{CO}$ obtained 
by means of the $CC$ ensemble applied to the ZGB model
and using a relatively small sample ($L\,=\, 32$). 
Starting from the stationary value of $\theta_{CO}$ at $P_{CO}
= 0.51$, one observes that stepwise increments of $\theta_{CO}$
cause $P_{CO}$ to increase steadily up to the $L$-dependent
upper spinodal point $P_{CO}^{S}(L\,=\,32) = 0.5330\,\pm\,0.031$. 
Turning around the spinodal point, further increments 
of $\theta_{CO}$ cause  $P_{CO}$ to 
decrease steadily up to $P_{CO} \approx\,0.513\,$ for
$\theta_{CO}^{max}\,\approx\,0.825$. 
At this point the growing $CO$ branch finishes and the subsequent
decrease in $\theta_{CO}$ causes the system to return to the starting
point where the decreasing $CO$ branch of the loop ends. 
Notice that both the growing and decreasing branches 
describe the same trajectory (within error bars) 
on the ($\theta_{CO}\,,\,P_{CO}$) plane.
So, it has been concluded that hysteretic effects are not observed 
using this small lattice \cite{weJPA}. 
\begin{figure}
\begin{center}
\epsfxsize=6cm
\epsfysize=6cm
\epsfbox{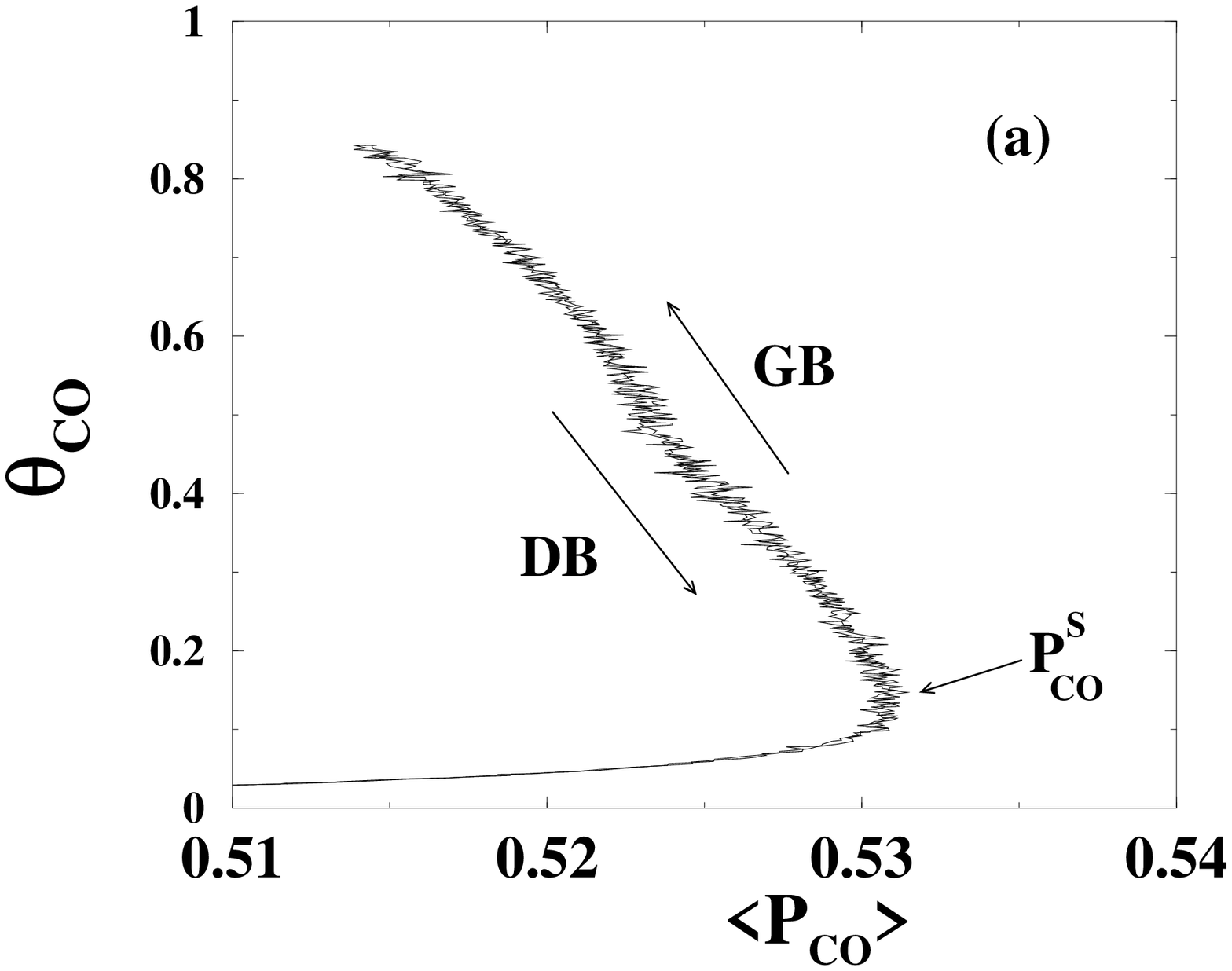}
\end{center}
\begin{center}
\epsfxsize=6cm
\epsfysize=6cm
\epsfbox{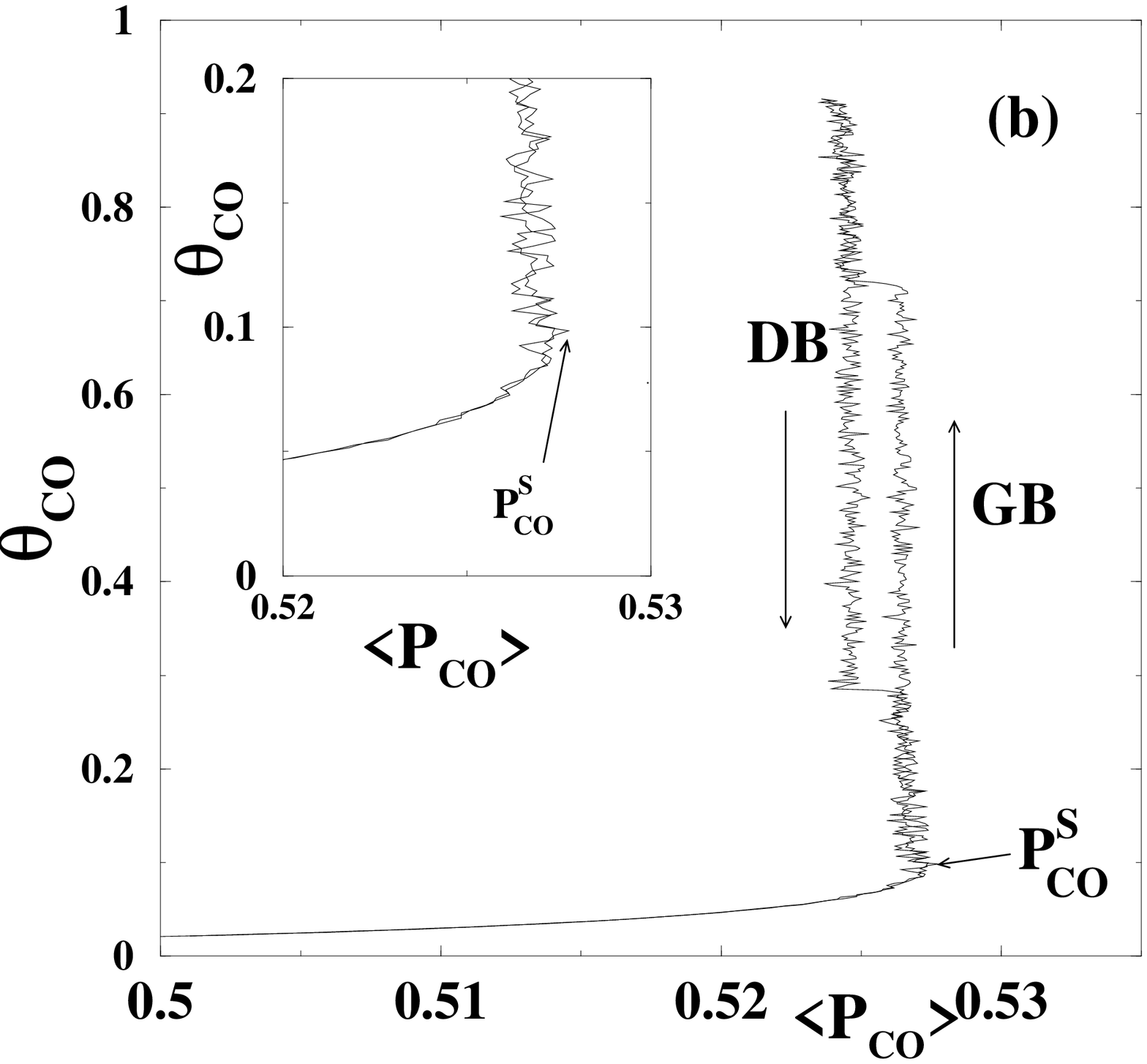}
\end{center}
\begin{center}
\epsfxsize=6cm
\epsfysize=6cm
\epsfbox{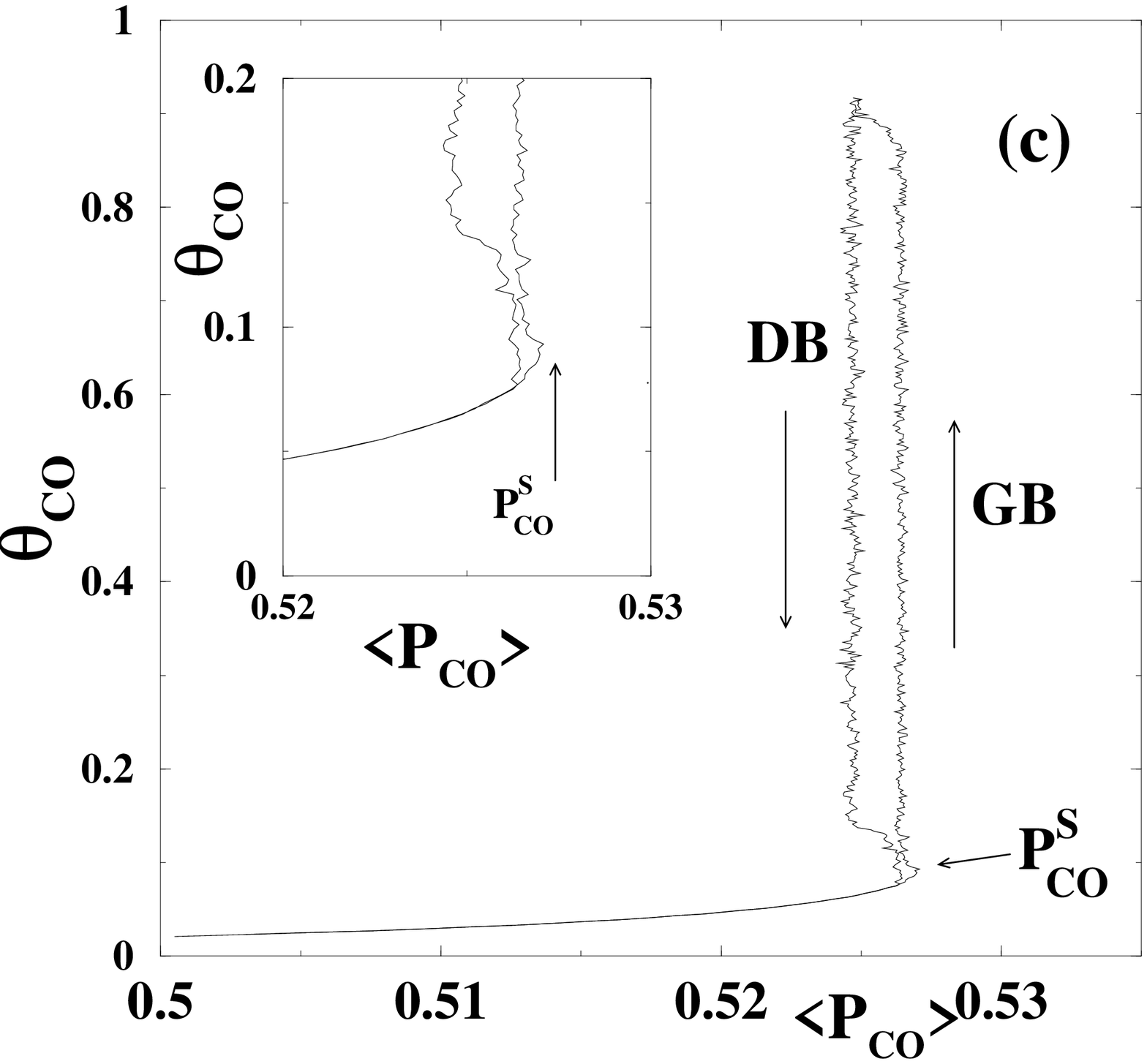}
\end{center}
\caption {Plots of $\theta_{CO}$ versus $<P_{CO}>$ obtained using
the constant coverage ensemble. Results corresponding to
$\tau_{P}\,=\,100\,$ mcs, $\tau_{M}\,=\,2000\,$  mcs and a stepwise
variation of $\theta_{CO}$ given by ${\Delta}{\theta}_{CO} \,= \, 2$ x
$10^{-3}$. The arrow pointing up (down) shows the growing (decreasing)
$\theta_{CO}$ branches of the loop. The upper spinodal point $
P_{CO}^{S}$ is also shown; a) $L\,=\,32$ ; b) $L\,=\,256$ and c)
$L\,=1024$. The insets of figures (b) and (c) show zooms of the
spinodal region.}
\label{Figure4}
\end{figure} 

Increasing the size of the lattice ($L\,=\,256$ in figure 4(b)), 
the behaviour of the system changes dramatically. 
On the one hand, the spinodal point becomes appreciably shifted 
(see inset of figure 4(b)), and  
on the other hand, hysteretic effects 
become evident since $CO$-growing and $CO$-decreasing branches can be
clearly distinguished.
Notice that within a remarkably wide range of $CO$ values both
branches are vertical, and consequently parallel each to other. 
 After increasing the lattice size ($L\,=\,1024$ in figure 4(c)) only minor 
changes in $P_{CO}^{S}$, $P_{CO}^{GB}$, $P_{CO}^{DB}$ occur, but 
a well defined spinodal point and a hysteresis loop can still be observed 
(see inset of figure 4(c)).
 
Figure 5 shows a plot of the $L$-dependent spinodal points 
($P_{CO}^{S}( L)$) versus the inverse lattice size
($L^{-1}$). Performing an extrapolation to the infinite size limit yields
$P_{CO}^{S}(L\,=\,\infty)\,=\,0.5270\,\pm\,0.0005$ \cite{weJPA}. This
figure should be compared with the value reported by Brosilow \etal
\cite{zb}, $P_{CO}^{S} \approx 0.5285$, which
corresponds to a determination of $P_{CO}^{S}$ in a finite
lattice of size $L = 1024$. Also, 
Evans \etal \cite{jim} have reported $P_{CO}^{S}\approx 0.527$ for a
finite lattice of size $L\,= 256$. Very recently, an independent 
estimation given by $P_{CO}^{S} = 0.52675 \pm 0.00025$, which is in excellent 
agreement with the figures measured using the CC method, has been 
obtained by means of short-time dynamic measurements \cite{pla}.  

Plots of both $P_{CO}^{GB}$ and $P_{CO}^{DB}$ versus
${L}^{-1}$, also shown in figure 5, indicate that finite-size
effects are negligible for large lattices (${L}\,>\,512$). 
So, the extrapolated estimates are \cite{weJPA} 
$P_{CO}^{GB}\,\cong\,0.5264\,\pm\,0.0002$ and
$P_{CO}^{DB}\,\cong\,0.52467\,\pm\,0.0003$, respectively.

\begin{figure}
\begin{center}
\epsfxsize=6cm
\epsfysize=6cm
\epsfbox{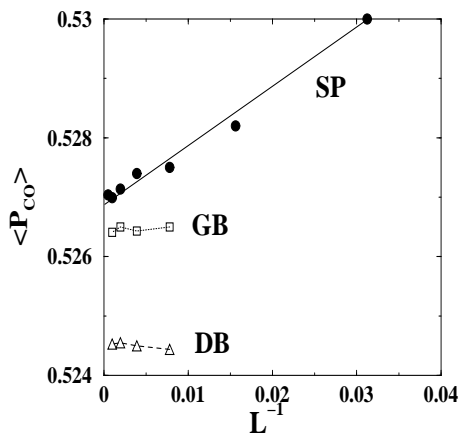}
\end{center}
\caption{Plots of the upper spinodal points determined using lattices
of different sizes ($\bullet \, P_{CO}^{S}(L)$), the
$CO$ pressure for the growing ($\Box \, P_{CO}^{GB}$), and
decreasing branches ($\bigtriangleup \, P_{CO}^{DB}$) versus the
inverse lattice size ($L^{-1}$). More details in the text.}
\label{Figure5}
\end{figure}

Another interesting feature of $CC$ simulations showing
hysteresis is that one can exchange the role of
the axis in the following sense: 
finely tuning the coverage $\theta_{CO}$ one can induce 
the system to undergo first-order transitions  
in parameter space ($P_{CO}$ in this case).

Hysteretic effects can be further understood after examination of 
various snapshot configurations as shown in figure 6. 
It should be noticed that all of the configurations belong to 
the coexistence region and, consequently, these states are not
allowed when simulating the ZGB model using the standard algorithm since 
right at $P_{2\,CO}$, $\theta_{CO}$ displays a discontinuous jump from 
$\theta_{CO}\,\approx\,10^{-3}$ to $\theta_{CO}\,=\,1$ (see figure 1). 
Figure 6(a) shows a typical configuration corresponding to the 
spinodal point $P_{CO}^{S} \simeq 0.5270$ with  
$\theta_{CO}\,\cong\,0.087\,$. Here, one observes that some small but
compact $CO$ clusters have already been nucleated. This configuration is
different from those obtained within the reactive regime using the
standard algorithm (not shown here for the sake of space) that show
mostly $CO$ monomers with $\theta_{CO}\,\approx\,10^{-3}$. 
The snapshot configurations shown in figures 6(b) and (c) correspond
to the growing branch and have been obtained for
$\theta_{CO}\,\approx\,0.30$ and $\theta_{CO}\,\approx\, 0.50$,
respectively. It should be noticed that above $P_{CO}^{S}$ a
single massive  $CO$ cluster has spread within the reactive phase. In
figure 6(b), this massive $CO$ cluster does not percolate, but
increasing $\theta_{CO}$ percolation along a single direction 
of the lattice is observed (figure 6(c)). Percolation of the massive 
cluster along only
one direction is observed up to a relatively high $CO$ coverage
($\theta_{CO}\,\cong\,0.763$ in figure 6(d)). These values of the
$\theta_{CO}$ are remarkably greater than the percolation threshold of
random percolation model, given by $P_{C}\,\approx\,0.59275$
\cite{stau}. However, the random percolation cluster is a fractal
object with fractal dimension ${\cal D}_{F}\,\cong\,1.89...$ \cite{stau}, while
the $CO$ cluster is a compact object. 
Dangling ends emerging from the surface of the $CO$ cluster eventually get in
contact causing percolation of such a cluster in both directions of the
lattice (figure 6(e)). It should be noticed that the snapshot
configuration of figure 6(d) corresponds to the growing branch while
that of figure 6(e), which has been taken after few mcs, corresponds to
an effective $CO$ pressure characteristic of the decreasing
branch. Therefore, the jump from one
branch to the other seems to be accompanied by a change in the
configuration of the $CO$ cluster. 
It will be interesting to quantitatively study the properties of the 
interface between  the $CO$ cluster and the
reactive phase in order to determine their possible self-affine nature,
as well as the interplay between curvature and hysteresis. 
From the qualitative point of view, the examination of snapshots
suggests that both the interface roughness and length of the massive 
$CO$ cluster remain almost unchanged for the growing branch. 
When $\theta_{CO}$ is further increased, the jump to 
the decreasing branch is eventually characterized by the
onset of percolation along both directions of the lattice and,
consequently, the macroscopic length and the curvature of the interface
may change. So, the subtle interplay of interfacial properties, such as length,
roughness, curvature, etc. has to be studied in detail in order to
fully understand the hysteresis loop observed using the $CC$ ensemble. 
It is expected that the longer the interface length of the growing
branch the easier the $CO-O$ reaction, so one needs a
higher effective $CO$ pressure to keep the coverage constant. 
In contrast, along the decreasing  
branch, the shorter length of the interface inhibits reactions so that
a greater oxygen pressure (smaller $CO$ pressure) is needed to achieve
the desired $CO$ coverage. So, these arguments may explain the
existence of two branches. 

\begin{figure}
\vskip 0.50 true cm
\begin{minipage}{5.0cm}
\begin{center}
\includegraphics[width=5.0cm]{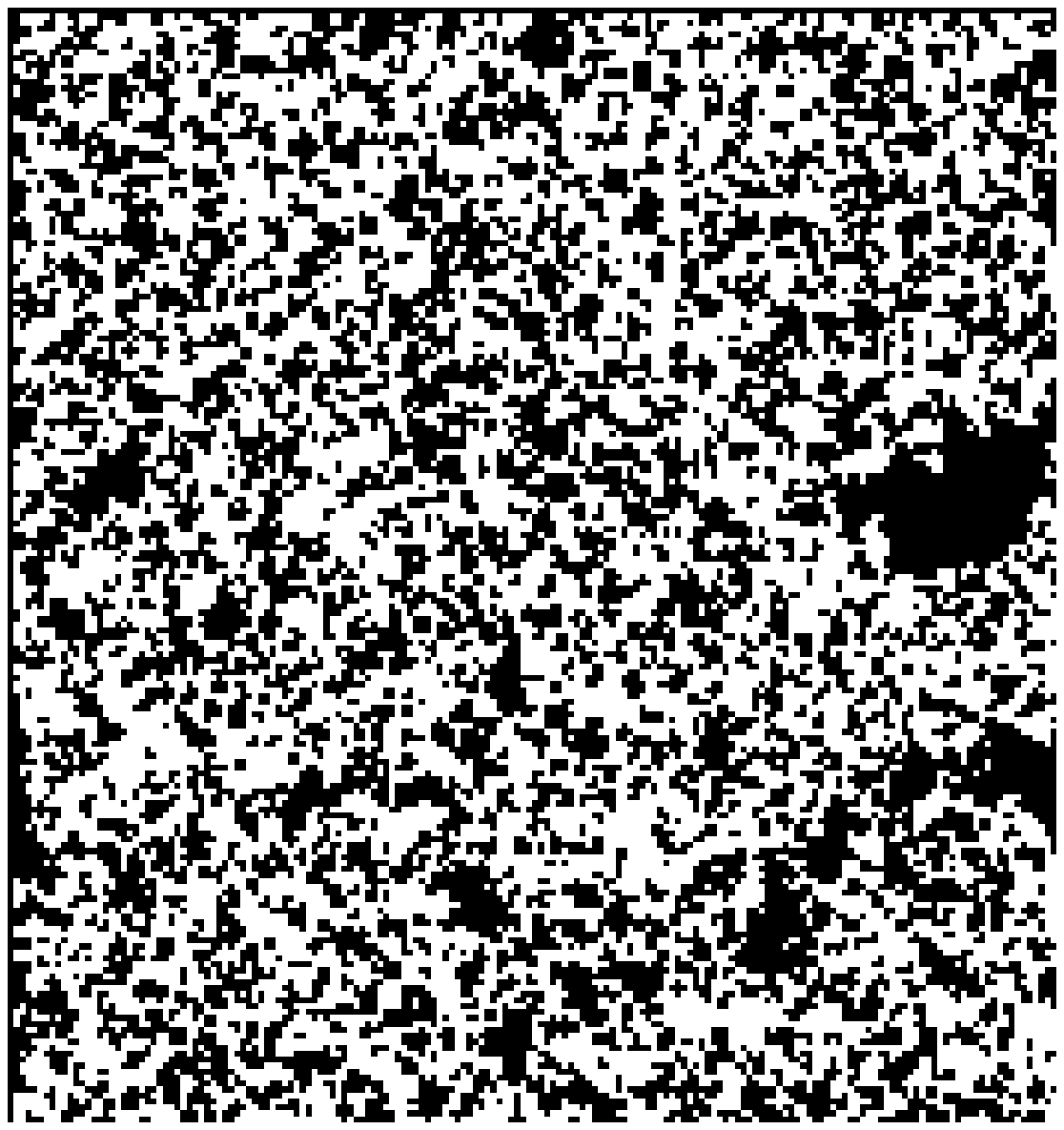}
\end{center}
\end{minipage}
\begin{minipage}{5.0cm}
\begin{center}
\includegraphics[width=5.0cm]{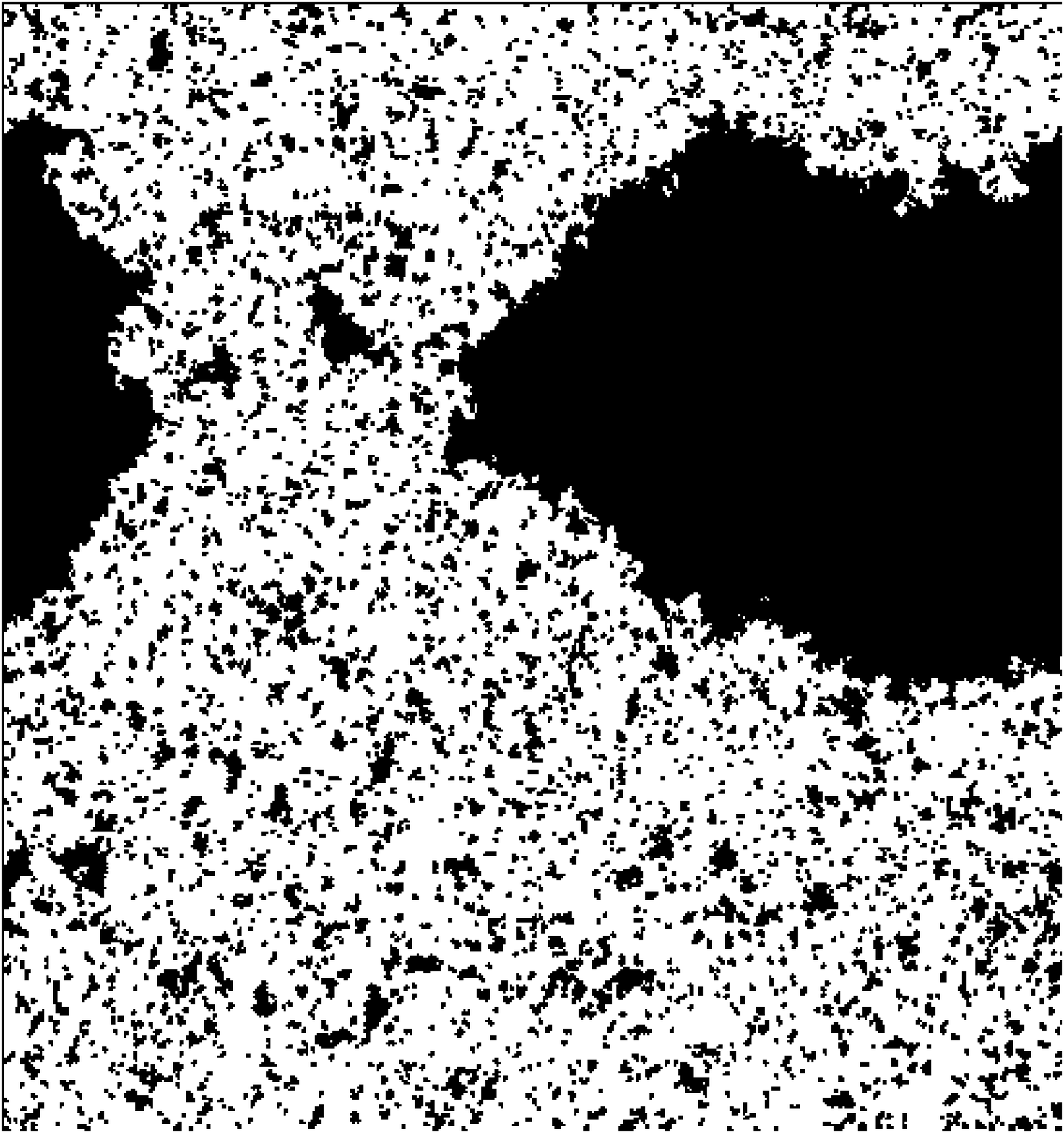}
\end{center}
\end{minipage}
\begin{minipage}{5.0cm}
\begin{center}
\includegraphics[width=5.0cm]{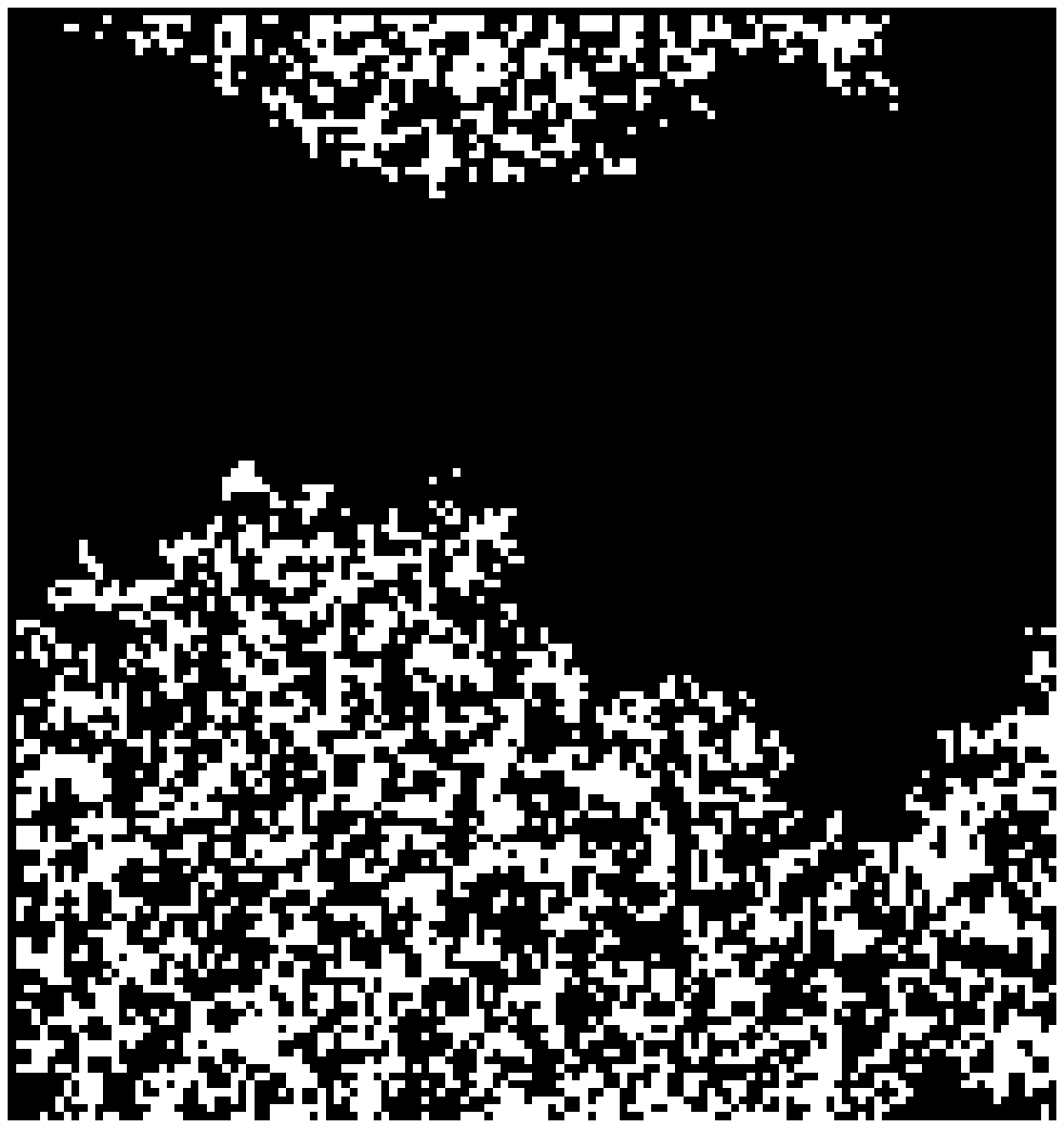}
\end{center}
\end{minipage}
\vskip 1.0 true cm
\begin{minipage}{5.0cm}
\begin{center}
\includegraphics[width=5.0cm]{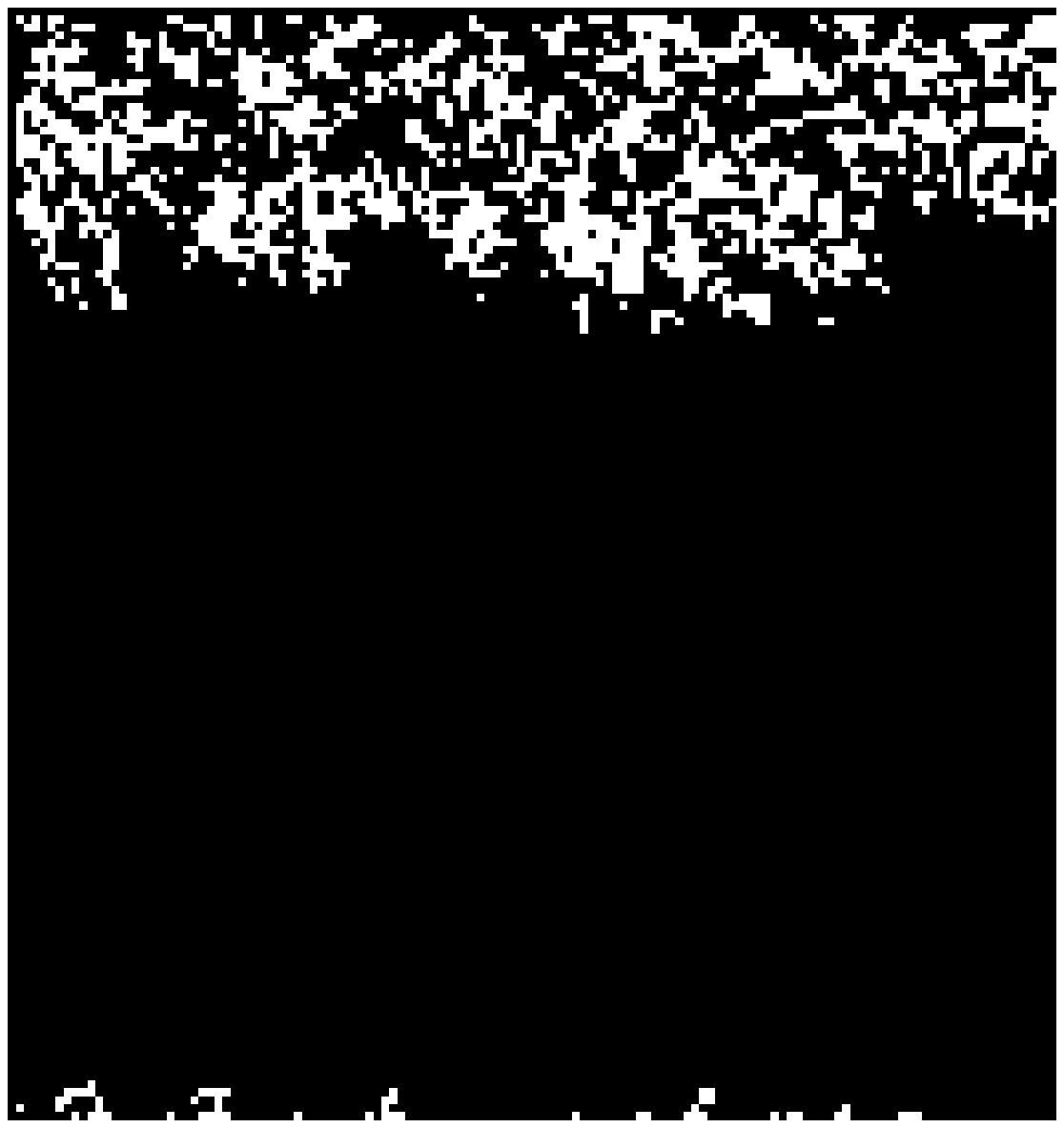}
\end{center}
\end{minipage}
\begin{minipage}{5.0cm}
\begin{center}
\includegraphics[width=5.0cm]{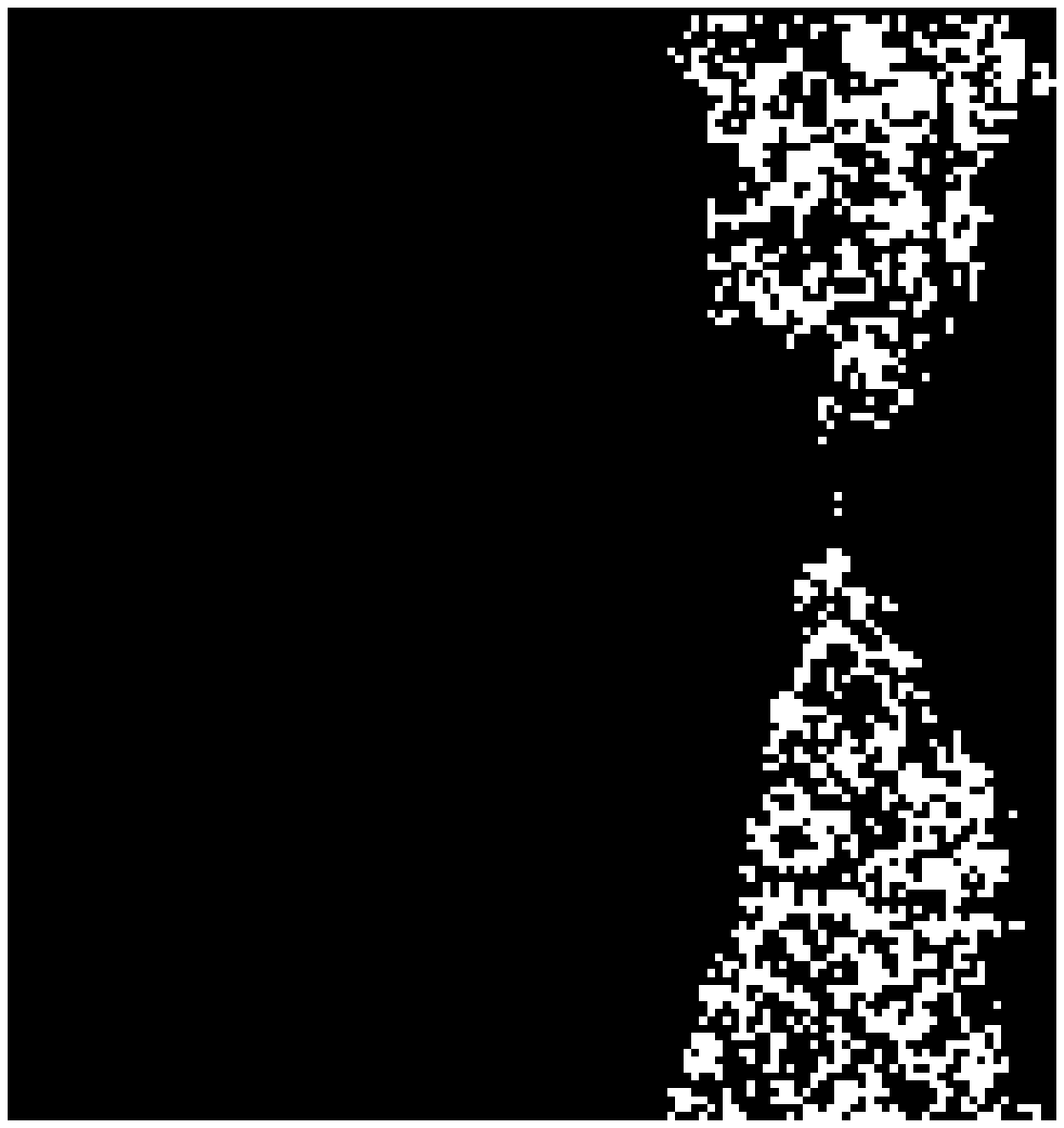}
\end{center}
\end{minipage}
\caption{Typical snapshot configurations obtained using the constant
coverage ensemble and lattices of side $L\,=\,512$, $CO$-occupied sites
are black while other sites are left white; a) snapshot obtained at
the spinodal point with $\theta_{CO}\,\approx\,0.087$, b), c) and d)
are snapshots obtained along the growing branch with
$\theta_{CO}\,\approx\,0.30$, $\theta_{CO}\,\approx\,0.50$, and
$\theta_{CO}\,\approx\,0.763$, respectively; e) snapshot obtained few
Monte Carlo steps after figure d) but now the system has
jumped to the decreasing branch with $\theta_{CO}\,\simeq\,0.78$.}
\label{Figure6}
\end{figure}

The snapshot configurations of figure 5 unambiguously show the
coexistence of two phases, namely a $CO$-rich phase dominated by a
massive $CO$ cluster that corresponds to the $CO$-poisoned state and
a reactive phase decorated with small $CO$ islands. It should be
noticed that such a nice coexistence picture is only available using
the $CC$ ensemble since coexistence configurations are not accessible
using the standard ensemble. It should also be noted that the 
existence of hysteretic effects hinders the location of the
coexistence point using the $CC$ ensemble method. In fact, in the 
case of hysteresis in thermodynamic equilibrium the chemical 
potential at coexistence can
be obtained after a proper thermodynamic integration of the growing and
decreasing branches of the hysteresis loop \cite{albin}. 
For nonequilibrium systems like the ZGB model
where no energetic interactions are considered, the standard methods of
equilibrium thermodynamics are not useful.  
In order to overcome this shortcoming a method based on the 
spontaneous creation algorithm already used to study different 
systems \cite{ront,bocca}, has been proposed \cite{weJPA}.
The method implies the study of the stability of the
hysteresis branches upon the application of a small perturbation. This
can be carried out by introducing a negligible small $CO$ desorption 
probability ($P_{CO}^{D}$) to the ZGB model. It is well
known that the first-order nature of the IPT of the ZGB model remains
if $P_{CO}^{D}\,<\,0.1$ \cite{albDES,zbDES}. It has been found that
taking $P_{D}\,=\,10^{-6}$ both branches
of the hysteresis loop collapse into a single one that can be identified as
the coexistence point given by $P_{2\,CO}\, \cong \, 0.52583(9)$ \cite{weJPA}. 
This figure is close to the middle of the hysteresis loop 
located close to $P_{2\,CO}\,\cong\,0.52554\,\pm\,0.00015$, 
where the error bars cover both branches of the loop. 
The value $P_{2\,CO}\,\approx\, 0.52560(1)$ reported by Brosilow \etal 
 \cite{zb}, which is remarkably close to this figure, has been
obtained using the $CC$ ensemble but neglecting both finite-size and
hysteretic effects. Regrettably, the size of the lattice used in
reference \cite{zb} was not reported, 
and therefore additional comparisons cannot be performed.

In the
seminal paper of Ziff \etal \cite{zgb}, an estimation of the coexistence
point is performed studying the stability of the coexistence phase.  
This analysis gives  $P_{2\,CO}\,\cong \,0.525(1)$, which is also in
good agreement with other independent estimates.
Also, Evans \etal \cite{jim} have reported  $P_{2\,CO} = 0.525 (1)$ based
on the analysis of epidemic simulations. The value reported 
by Meakin \etal \cite{meak}, $P_{2\,CO} = 0.5277$, seems to be 
influenced by metastabilities
due to the large lattices used for the standard simulation method.  
Therefore, that figure is a bit larger and may correspond to a value
close to the spinodal point. Surprisingly, better values, e.g. 
$P_{2\,CO} \; 0.5255(5)$ \cite{paki}, 
can be obtained using very small lattices and the
standard algorithm since in such samples metastabilities are short-lived. 

Very recently, hysteresis phenomena have been studied on the basis 
of a modified version of the ZGB model \cite{hua}.
In fact, it is assumed that the surface of the catalyst has
two kinds of inhomogeneities or defects. In type-1 defects, which are
randomly distributed on the sample with probability $p_{1}$, the 
desorption of adsorbed $CO$ proceeds with  probability $p_{des1}$.
In type-2 inhomogeneities, which are randomly distributed with
probability $p_{2}$, the adsorption of oxygen molecules
is inhibited and the desorption probability of $CO$ is given by $P_{des2}$.
Furthermore, $p_{des2} < p_{des} < p_{des1}$, where $p_{des}$ is the
desorption probability of $CO$ on unperturbed lattice sites.
Also, the diffusion of $CO$ species is considered with probability 
$p_{diff}$, while the probability of other events such as
adsorption, desorption and reaction is given by $p_{chem} = 1 - p_{diff}$.
In order to study hyteretic effects the $CO$ pressure is
varied at a constant rate in closed cycles.
It is interesting to discus the mechanisms, which are 
originated by the definition of the model, which may lead to 
hysteretic effects. In fact, the low desorption
probability of $CO$ prevents the occurrence of a $CO$ absorbing state
and, consequently, the abrupt transition from the high-reactivity state
to the low-reactivity regime is reversible. Furthermore, the 
blocking of lattice sites for oxygen adsorption also prevents the
formation of the oxygen-poisoned state and the second-order
IPT becomes reversible and takes place between an oxygen-rich
low-reactivity regime and a high-reactivity state. Since escaping 
from both low-reactivity regimes is quite difficult, the occurrence of
hysteresis in dynamic measurements can be anticipated.   
 
The reported results obtained by means of simulations
\cite{hua} (see figure 7) are in qualitative agreement 
with the experimental findings (figure 3). Figure 7(a)
corresponds to the evolution of oxygen coverage. $\theta_{O}$ 
decreases smoothly from the oxygen-rich state when the $CO$ pressure 
is raised. This behaviour resembles the case of the ZGB model
(figure 1). However, the abrupt increase in $\theta_{O}$ observed 
when $P_{CO}$ is lowered is due to the fact that the surface is
mostly covered by $CO$ species (see figure  7(b)), which have low 
desorption probability. In fact, oxygen can only be adsorbed 
on a nearest-neighbor pair of vacant sites that may become
available with (roughly) a low  probability of the order of  
$(1-P_{CO}) p_{des}^2$. On the other hand, the growing
branch of $\theta_{CO}$ (figure 7(b)) is flat and that coverage 
remains close to
zero, for $P_{CO} < 0.48$. Subsequently, it exhibits an abrupt increase 
close to $P_{CO} \simeq 0.5$, as already expected from the
behaviour of the ZGB model (figure 1). The high coverage
region of the decreasing branch is due to the 
low desorption probability of $CO$ while the subsequent
sudden drop is consistent with the abrupt increase in 
$\theta_{O}$ (figure 7(b)). It should be noticed that
the experimental measurement of the photocurrent
does not allow one to distinguish $\theta_{CO}$ from 
$\theta_{O}$ \cite{hister}, so figure 3(a) should be 
compared with a composition of both figures 7(a) and 7(b). Anyway, the
experimental findings are nicely (qualitatively) 
reproduced by the simulations. Finally, the behaviour of the 
rate of $CO_{2}$ production (figure 7(c) is also in excellent 
agreement with the experimental data shown in figure 3(b).
\begin{figure}
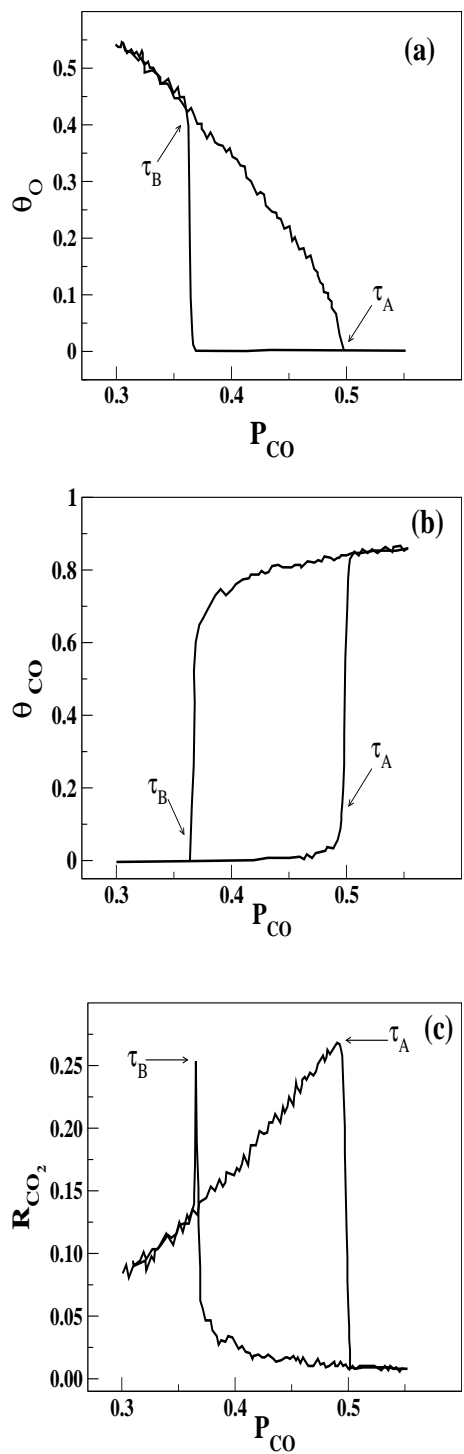

\begin{center}
\epsfxsize=6cm
\epsfysize=6cm
\epsfbox{Fig7a.eps}
\end{center}
\begin{center}
\epsfxsize=6cm
\epsfysize=6cm
\epsfbox{Fig7b.eps}
\end{center}
\vskip 0.5 true cm
\begin{center}
\epsfxsize=6cm
\epsfysize=6cm
\epsfbox{Fig7c.eps}
\end{center}
\caption{Simulation results of hysteresis phenomena obtained 
using a modified version of the ZGB model, according to 
Hua and Ma \cite{hua}. The simulation parameters are: 
$p_{chem} = 0.01$, $p_{des} = 0.1$, $p_{des1} = 0.8$, $p_{des2} = 0.05$,
and $p_{1} = p_{2} = 0.1$. The scanning rate of $P_{CO}$
is $\beta_{CO} = \frac{dP_{CO}}{dt} = 0.002/(10 mcs)$.
(a) and (b) correspond to the coverages while (c)
shows the rate of $CO_{2}$ production. More details in the 
text. Adapted from reference \cite{hua}.}    
\label{Figure7}
\end{figure} 
As in the experiments \cite{hister} (see figure 3), 
the numerical simulation results shown in figure 7
also allow one to estimate the values of the $CO$ partial pressure
where the transitions from the monostable states A and B
to the bistable state, $\tau_{B}$ and $\tau_{A}$, respectively
take place. Therefore, the width of the hysteresis loop is given by
$\Delta \tau = \tau_{A} - \tau_{B}$. The dependence of
$\Delta \tau$ on the scan rate of the $CO$ pressure
($\beta_{CO} = \frac{dP_{CO}}{dt}$) has also been
measured experimentally and by means of simulations, as
shown in figures 8(a) and 8(b), respectively.  
\begin{figure}
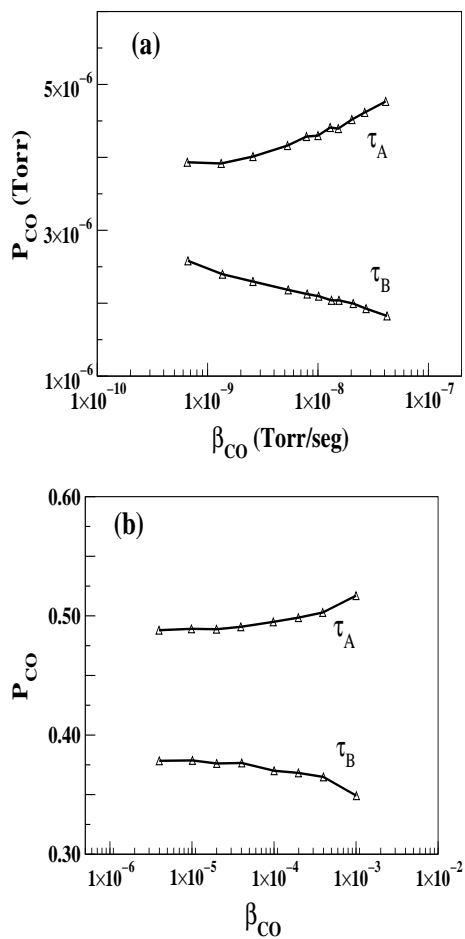

\begin{center}
\epsfxsize=6cm
\epsfysize=6cm
\epsfbox{Fig8a.eps}
\end{center}
\begin{center}
\epsfxsize=6cm
\epsfysize=6cm
\epsfbox{Fig8b.eps}
\end{center}
\caption{Plots of the $CO$ partial pressure
where the transitions from the monostable states A and B
to the bistable state, $\tau_{B}$ and $\tau_{A}$, respectively
take place; versus the scanning rate of $P_{CO}$ given by 
$\beta_{CO} = \frac{dP_{CO}}{dt}$
(a) Experimental data measured under the same conditions as
those shown in figure 3. Adapted from reference \cite{hister}.
(b) Simulation results obtained using the parameters listed
in figure 7 but varying the scanning rate of $P_{CO}$ given
in units of $mcs^{-1}$. Adapted from reference \cite{hua}.}  
\label{Figure8}
\end{figure}
Again, the numerical results are in remarkable qualitative agreement
with the experiments. Neglecting surface defects and based on 
mean-field calculations Zhdanov and Kasemo \cite{rev3} have 
stated that the two branches of the
hysteresis loop should fall together to the equistability point,
provided there occurs a sufficiently rapid nucleation and growth 
of islands. This result is expected to be valid on the 
limit $\beta_{CO} \rightarrow 0$ where one also should observe
$\Delta \tau \rightarrow 0$. As shown in figure 8, the transition 
points $\tau_{A}$ and $\tau_{B}$ approach each other and 
$\Delta \tau$ shrinks with decreasing $\beta_{CO}$. 
Conclusive evidence on the validity of the mean-field prediction
cannot be found due to either experimental or numerical
limitations to achieve a vanishing small scanning rate. However,
it has been suggested that $\delta \tau$ should be finite due to the
presence of surface defects \cite{hister,hua}, which is neglected
in the mean-field treatment \cite{rev3}.   
 
Comparing the numerical results obtained applying the 
CC method to the ZGB model \cite{weJPA} with those of the study
performed by Hua \etal \cite{hua}, one has to consider that not only  
both models are different, but also  
the data corresponding to the CC ensemble were obtained after a 
long-time stabilization period and consequently exhibit smaller 
hysteretic effects, in contrast to the dynamic measurements
where the relaxation of the system towards stationary states is
not allowed \cite{hua}. 

As anticipated above a truly $CO-$poisoned state cannot 
be achieved in the experiments due to the nonvanishing
$CO-$desorption probability ($P_{des}^{CO}$). According to the
theory of thermal desorption and the experiments, $P_{des}^{CO}$
depends on the temperature of the catalysts and the energetic
interactions with neighboring adsorbed species through an
Arrhenius factor \cite{rh,cw}. Therefore, the magnitude of the 
abrupt drop in the reaction rate (see figure 2) decreases 
upon increasing $T$, as shown in figure 9 for the case of 
the catalytic oxidation of $CO$ on $Pt(111)$ \cite{bloc,bloc1}.
Furthermore, on increasing $T$ the sharp peak of the
reaction rate becomes rounded and for high enough
temperature, e.g. for $T \geq 524 K$ in figure 8, 
the signature of the first-order transition vanishes.
\begin{figure}
\begin{center}
\epsfxsize=6cm
\epsfysize=6cm
\epsfbox{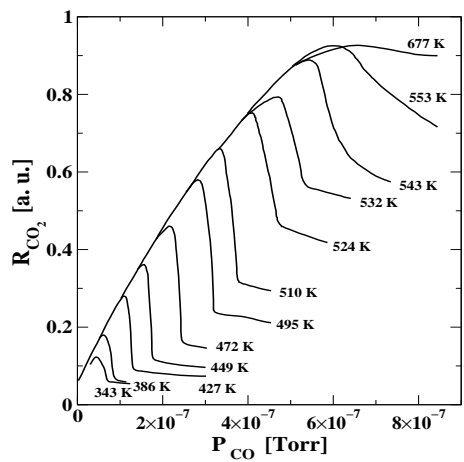}
\end{center}
\caption {Plots of the reaction rate of $CO_{2}$ production
versus the the partial pressure of $CO$. Experiments corresponding
to the catalytic oxidation of $CO$ on $Pt(111)$ single crystal, 
performed keeping the oxygen partial pressure constant 
($P_{O} = 2.0 \time 10^{-6}$ Torr) and varying the temperature
of the catalyst, as shown in the figure. 
Adapted from references \cite{bloc,bloc1}.} 
\label{Figure9}
\end{figure}

The influence of $CO$ desorption on the phase diagram of the 
ZGB model has also been studied by means of Monte Carlo 
simulations \cite{albDES,zbDES,hua}. The simplest approach is 
just to introduce an additional parameter to the ZGB model, 
given by the desorption probability $P_{des}^{CO}$. As expected, the
second-order IPT of the models is not influenced 
by $CO$ desorption \cite{albDES}.
However, the first-order IPT actually disappears because
due to the finite value of $P_{des}^{CO}$ the system can no longer achieve 
a truly $CO-$poisoned state. However, the first-order nature of the 
transition remains for very low desorption probabilities,
as shown in figure 10, for $P_{des}^{CO}\,<\,0.1$ \cite{albDES,zbDES}.
On increasing $P_{des}^{CO}$, the peak of the rate of $CO_{2}$ production 
becomes shifted and rounded in qualitative agreement with the 
experiments (figure 9). 

\begin{figure}
\vskip 1.0 true cm
\begin{center}
\epsfxsize=6cm
\epsfysize=6cm
\epsfbox{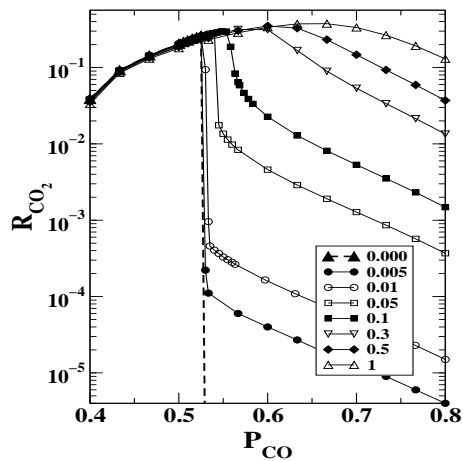}
\end{center}
\caption {Log-lineal plots of the reaction rate of $CO_{2}$ production
versus the the partial pressure of $CO$. Results obtained performing 
Monte Carlo simulations of the ZGB model for different desorption
probabilities ($P_{des}^{CO}$), as listed in the figure.} 
\label{Figure10}
\end{figure}

Another useful approach to the study of first-order IPT's
is to apply the EM as described in section \Sref{epidemias}.
Early epidemic studies of the first-order IPT of the
ZGB model have been performed by Evans and Miesch \cite{jim}.
Epidemic simulations were started with the surface of the 
catalysts fully covered by $CO$ species, except for an empty
patch placed at the center of the sample.
The time dependence of the number of empty sites ($N(t)$)
and the survival probability of the patches ($P(t)$)
were analyzed in terms of the conventional scaling 
relationships given by equations (\ref{nu}) and (\ref{so}).
An interesting feature observed in these earlier simulations
was the monotonic decrease of $N(t)$, which can be fitted
with an exponent $\eta \simeq -2.2$ \cite{jim}. This result
is in marked contrast with
the behaviour expected for second-order IPTs in the
DP universality class, where
equation (\ref{nu}) holds with a positive exponent
such as $N(t) \propto t^\eta$ with $\eta \approx 0.2295$ 
\cite{bobZ,muve} in two dimensions. Furthermore, it has been
observed that empty patches have a extremely low survival 
probability \cite{jim} and the data can be fitted using equation
(\ref{so}) with an exponent $\delta \simeq 1.97$, i.e a figure
much larger than the exponent expected for DP given by
$\delta \simeq 0.4505$ \cite{bobZ,muve}. 

Wide experience gained studying {\bf reversible} 
first-order critical phenomena shows that in this kind of
transitions the correlations are short-ranged \cite{gene}.
Therefore, the reported power-law decays of $N(t)$ 
and $P(t)$ \cite{jim} are certainly intriguing. 
However, recent extensive numerical simulations     
performed averaging results over $10^{9}$ different epidemic
runs have changed this scenario \cite{weJPA}, as shown 
in figure 11. In fact, data taken for $P_{CO}^{GB}$, $P_{2\,CO}$, 
and $P_{CO}^{DB}$ show pronounced curvature with a clear cut-off, 
departing from a power-law behaviour as described by equation (\ref{nu}). 
So, it has been concluded that the occurrence of power law
(scale invariance) in the first-order dynamic critical behaviour of the
ZGB model can safely be ruled out \cite{weJPA}.
On the other hand, for $P_{CO}\,\geq\,P_{CO}^{GB}$, log-log plots of
$N(t)$ versus $t$ exhibit pseudo power-law behaviour over
many decades ($10^{1} \leq t \leq 10^{3})$, as shown in figure 11.
The effective exponent describing the early time
behaviour of $N(t)$ is $\eta^{eff} \,\approx \,-2.0 \pm 0.1$,
in agreement with the result reported by Evans \etal \cite{jim}. 
However, after a long time, few successful epidemics prevail and 
the number of empty sites suddenly grows as $N(t)\propto t^{2}$, 
indicating a spatially homogeneous spreading.
\begin{figure}
\begin{center}
\epsfxsize=6cm
\epsfysize=6cm
\epsfbox{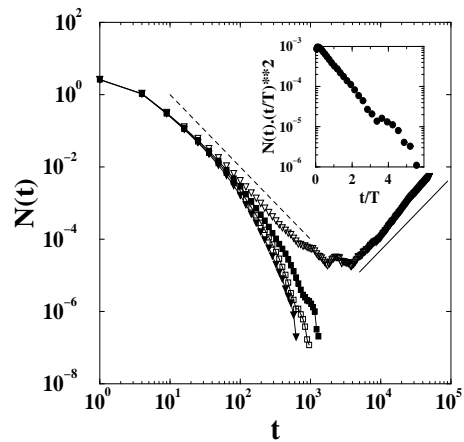}
\end{center}
\caption{Log-Log plots of the number of vacant sites $N(t)$ versus $t$ 
obtained performing epidemic studies of the ZGB model. 
Results averaged over $10^9$ different runs
($\blacktriangledown$ $P_{CO}^{GB}$,
$\blacksquare$ $P_{CO}^{DB}$,
$\square$ $P_{2CO}$, $\bigtriangledown$ $P_{CO} = 0.52345$).
For the latter, two straight lines have been drawn for the sake of
comparison: the dashed one with slope $\eta^{eff}= -2$ and the full one with
slope $2$. The inset shows a semilogarithmic plot of
$N(t)(T/t)^{-2}$ versus $t/T$ with $T=183$, according to equation (\ref{anz}),
obtained at $P_{2CO}$. More details in the text.}
\label{Figure11}
\end{figure}
The results shown in figure 11 suggest that 
instead of the power-law behaviour characteristic of second-order
transitions (equation (\ref{nu})), the epidemic behaviour close to
first-order IPT's could be described by means of a modified 
ansatz involving a short-time power-law behaviour followed 
by a long-time exponential decay, as given 
by equation (\ref{anz}) \cite{weJPA}. The inset of figure 11 shows 
a test of equation (\ref{anz}), namely a semilogarithmic plot of
$N(t) (\tau/t)^{2}$ versus $t/{\tau}$, where $\eta^{eff} = -2$ 
has been assumed. The scattering of points for long times is
simply due to the reduction of statistics as a consequence of the
low survival probability of the initial epidemic patches.
Summing up, epidemic studies of the ZGB model close to and at coexistence
show a pseudo power-law behaviour for short times ($t < T$)
that crosses over to an asymptotic exponential decay for longer times.
Consequently, the absence of scale invariance in the first-order
IPT of the ZGB model places this kind of transition on the same
footing its their reversible counterparts.

Another interesting scenario for the study of bistable behaviour
close to first-order IPT's, observed in catalyzed reactions,
is to investigate the properties of interfaces
generated during the propagation of reaction fronts. 
In fact, key experiments have underlined the
importance of front propagation for pattern formation in bistable
systems, including the formation of labyrinthine patterns \cite{lee1},
self-replicating spots \cite{lee2}, target patterns and spiral
waves \cite{zai,win}, stationary concentration
patterns (`Turing structures') \cite{tur}, etc.
Furthermore, recent experimental studies of catalytic
surface reactions have confirmed the existence of a wealth
of phenomena related to pattern formation upon
front propagation \cite{B,ron,ronert,xx1,xx2,dani,rose}.

The basic requirement for the observation of front propagation is
a process involving an unstable phase, which could be
displaced by a stable one, leading to the formation of an
interface where most reaction events take place.
This interesting situation is observed close to first-order
IPTs, as in the case of the ZGB model (figure 1).
In fact, just at the transition point $P_{2CO}$ one has a
discontinuity in $\theta _{CO}$ that corresponds to the
coexistence between a reactive state with small $CO$ clusters
and a $CO$-rich phase, which likely is a large $CO$-cluster,
as suggested by simulations performed using the CC ensemble
(see the snapshots of figures 6(b)-(e)).
Between $P_{2CO}$ and the upper-spinodal point $P_{CO}^{US}$,
the reactive state is unstable and it is displaced by the
$CO$-rich phase. On the contrary, between the lower spinodal
point $P_{CO}^{LS}$ and $P_{2CO}$ the reactive state displaces
the $CO$-rich phase. This latter case has been studied by
Evans and Ray \cite{jimn}, who have reported that the reactive regime displaces
the $CO$-poisoned state, resulting in a propagation
velocity $(V_p)$ normal to the interface. It has been proposed
that $V_p$ must vanish as $(P_{CO} \rightarrow P_{2CO})$ \cite{jimn},
where both states become equistable, so one has
\begin{equation}
V_p \propto (P_{CO} - P_{2CO})^{-\gamma} , 
\label{velo}
\end{equation}
\noindent with $\gamma >0$. The limit of high diffusivity of
the reactants can be well described by mean-field
reaction-diffusion equations, which give $\gamma =1$ \cite{jimn}.
It is interesting to notice that if diffusion is restricted or
even suppressed, simulation results give values of $\gamma $
that are also very close to unity, suggesting that this exponent
is independent of the surface diffusivity of the reactants \cite{jimn}.

For an evolving interface, there is a clear distinction between
the propagation direction and that perpendicular to it. So it may not be 
surprising that scaling is different along these two directions. 
Therefore, an interface lacks self-similarity but,
instead, can be regarded as a self-affine object \cite{bara}.
Based on general scaling arguments it can be shown that the
stochastic evolution of a driven interface along a strip of width $L$ is
characterized by long-wavelength fluctuations $(w(L,t))$ that have the
following time- and finite-size-behaviour \cite{bara}
\begin{equation}
w(L,t) \propto L^{\alpha} F(t/L^z) ,
\label{FV}
\end{equation}
\noindent where $F(x) \propto x^{\beta^{*}}$ for $x \ll 1$ and
$F(x) \rightarrow 1$ for $x \gg 1$, with $z = \alpha/\beta^{*}$.
So, the dynamic behaviour of the interface can be described in terms of
the exponents $\alpha$ and $\beta^{*}$, which are
the roughness and growth exponents, respectively. Thus, for an infinite
system $(L \rightarrow \infty)$, one has $w(t) \propto t^{\beta}$,
as $t \rightarrow \infty$. Note that $w$ is also known as the
interface width.

It is reasonable to expect that the scaling behaviour should
still hold after coarse-graining and passing to the
continuous limit. In fact, the dynamics of an interface
between two phases, one of which is growing into
the other, is believed to be correctly described by simple
nonlinear Langevin type equations, such as equation (\ref{kkppzz})
proposed by Kardar, Parisi and Zhang (KPZ) \cite{kpz}, the
Edward-Wilkinson (EW) equation \cite{EW}, and others \cite{bara}.

As in the case of second-order phase transitions, taking into
account the values of the dynamic exponents, evolving interfaces
can be grouped in sets of few universality classes, such that
interfaces characterized by the same exponents belong to
the same universality class. Among others, KPZ and EW universality
classes are the most frequently found in both
experiments and models \cite{bara}, including electrochemical
deposition, polycrystalline thin-film growth, fire-front
propagation, etc. \cite{damian,clar,evaff,ale}.

Pointing again our attention to the simulation results of
Evans and Ray \cite{jimn}, they have reported that
the propagation of the reaction interface, close to $P_{2CO}$,
can be described in terms of dynamic scaling arguments \cite{jimn},
with $\beta^{*} \simeq 0.3$, i.e., a figure close
to the KPZ value ($\beta^{*} = 1/3$ in $d=2$ dimensions).

Very recently, Ch\'avez \etal \cite{cha} studied
the dynamics of front propagation in the catalytic oxidation of CO
on $Pt(100)$ by means of a cellular automaton simulation. It is found that
the dynamic scaling exponents of the interface are well described
by equation (\ref{FV}) with $\alpha = 1/2$ and $\beta^{*} = 1/3$.
It is also reported that, in the absence of surface diffusion, the
interface dynamics exhibits KPZ behaviour \cite{cha}.

Based on a variant of the ZGB model, Goodman \etal \cite{dios}
have studied the propagation of concentration waves. They reported
the observation of trigger waves within the bistable regime of
the process, i.e., close to the first-order IPT. In fact,
within this regime one has the coexistence of a stable state
with a metastable one. At the boundary between the two, the stable state
will displace the metastable one and the boundary will move, so this process
leads to the propagation of concentration fronts (trigger waves).
Goodman \etal \cite{dios} found that the velocity of the
$CO$ front depends on the diffusion rate $D_{CO}$ of $CO$ species
(diffusion of oxygen is neglected) and $(P_{CO})$. The velocity of
the front vanishes on approaching the poisoning transition
at $P_{2CO}(D_{CO})$ (note that the transition point now
depends on $D_{CO}$), according to equation (\ref{velo}), with $\gamma
\simeq 1$, in agreement with the results of Evans \etal \cite{jimn}.

While front propagation during the catalytic oxidation
of $CO$ on platinum surfaces has been observed in numerous
experiments \cite{B,xx2,ron,ronert,rose}, the quantitative analysis
of data is somewhat restricted by the fact that the
fronts are neither flat nor uniformly curved, eventually
several of them nucleate almost at the same time and, in
most cases, the occurrence of strong interactions between
fronts (`interference of chemical waves') makes clean
interpretations quite difficult. In order to overcome these
shortcomings, Haas \etal \cite{xx1} have studied the
propagation of reaction fronts on narrow channels
with typical widths of 7, 14 and 28 $\mu m$. 
The main advantage of these controlled quasi-one-dimensional
geometries is that frequently only a single front propagates
along the channel, thus avoiding front interactions.
Additionally, the narrowness of the channels and the absence
of undesired flux at the boundaries lead to practically
planar fronts. Using this experimental setup, the
front velocity in different channels can be measured as
a function of the partial pressure of $CO$, keeping the
temperature and the oxygen partial pressure constant, as shown in
figure 12(a). At low $P_{CO}$ values only oxygen fronts are
observed. Furthermore, their velocity decreases when $P_{CO}$
is increased, reaching a minimum value at a certain
critical threshold $P_{CO}^{crit2}$ (see figure 12(a)).
When $P_{CO}$ is further increased a jump is observed- now
the front reverses itself into a $CO$ front and travels
in the opposite direction. When $P_{CO}$ is lowered from high
values, the $CO$ fronts become slower and hysteresis
is observed (see the coexistence between Oxygen and $CO$ fronts
in figure 12(a) for $P_{CO} < P_{CO}^{crit2}$). Finally,
at $P_{CO}^{crit1}$ another jump is observed- under these
conditions $CO$ fronts can no longer persist below a quite
low (but nonzero velocity) and they reverse themselves into fast
Oxygen fronts (figure 12(a)).
\begin{figure}
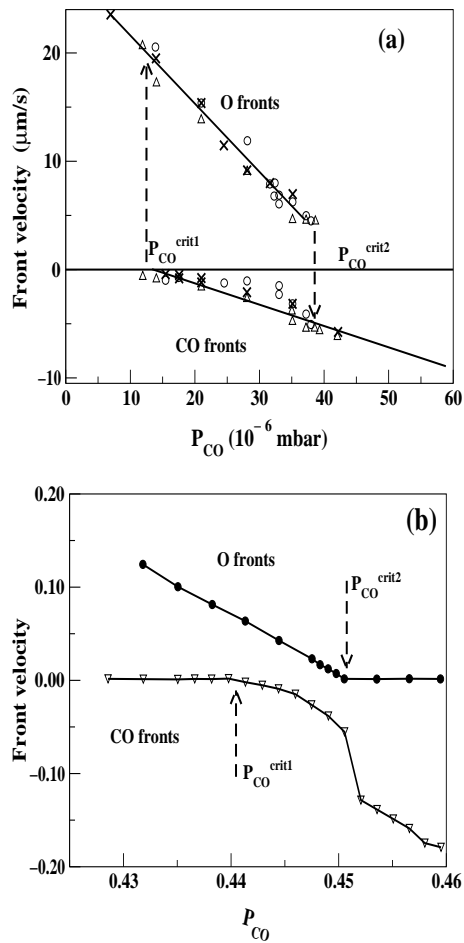

\begin{center}
\epsfxsize=6cm
\epsfysize=6cm
\epsfbox{Fig12a.eps}
\end{center}
\begin{center}
\epsfxsize=6cm
\epsfysize=6cm
\epsfbox{Fig12b.eps}
\end{center}
\caption{(a) Experimental results showing the front velocity inside 
three channels (triangles: $28 \mu m$, circles $14 \mu m$, and
crosses $7 \mu m$) as a function of $CO$ partial pressure.
Data measured taking $P_{O} = 4 \times 10^{-4}$ mbar and 
$T = 360$ K. More details in the text. Adapted from reference \cite{xx1}.
(b) Monte Carlo results obtained with the ZGB model for front 
propagation in channels of $L = 10$ lattice units (LU). Plots of 
the front velocity (in units of LU per mcs) versus $CO$ partial 
pressure. The lines show the critical pressures at which propagation
stops. Adapted from reference \cite{evawaves}. }  
\label{Figure12}
\end{figure}
Many features of the experiment of Haas \etal \cite{xx1}
can be recovered simulating front propagation with the aid of the
ZGB model on the square lattice with rectangular geometries of
sides $L \times M$ ($L\ll M$) \cite{evawaves}. Thus $L$ is the width of the
channel and $M$ its length. Free boundary conditions were taken
along the channel while the opposite ends are assumed to be in contact
with oxygen and $CO$ sources, respectively. If $O$ or $CO$ species
are removed from the ends of the channels (i.e., the `sources'),
due to the reaction process, they are immediately replaced. The
propagation of the $CO$ concentration profile
was studied starting with a sample fully covered by oxygen, except
for the first and second columns, which are covered by $CO$
(the $CO$ source), and left empty, respectively. The propagation of
the oxygen profile was followed using a similar procedure \cite{evawaves}.
Under these conditions one always has two competing
interfaces along the channel.

In order to make a quantitative description of the propagation, 
the concentration profiles of the reactants, $\theta_{O}(x)$ and
$\theta_{CO}(x)$, are measured along the length of the channel $x$
in the $M$-direction and averaged over each column of lattice
sites of length $L$. Then, the moments of the profiles,
which in subsequent steps can be used to determine the propagation
velocity and the width of the profiles, are also measured. In fact,
the moments of $n$th order of the profiles can be evaluated
according to \cite{kan1}
\begin{equation}
<x^n>_{\theta} = \frac{\sum x^n [\theta(x+1) -
\theta(x)]}{\sum [\theta(x+1) - \theta(x)]} .
\label{moment}
\end{equation}
\noindent Thus, using equation (\ref{moment}) the velocity of
propagation can be obtained from the first moment
\begin{equation}
V =  \frac{d<x>}{dt} . 
\label{velocity}
\end{equation}

Monte Carlo simulation results show that the front propagation
velocity depends on both $P_{CO}$ and the channel width $L$,
as shown in figure 12(b). This figure also shows that the
displacement of $CO$- and $O$-poisoned channels by the reactive
regime stops at certain ($L$-dependent) critical values,
$P_{CO}^{c2}(L)$ and $P_{CO}^{c1}(L)$, respectively.  By means of an
extrapolation to the thermodynamic limit it is possible  to identify
these critical values with the critical points of
the ZGB model, namely  $P_{CO}^{c1}(L \rightarrow \infty) = P_{1CO}$
and  $P_{CO}^{c2}(L\rightarrow \infty) = P_{2CO} $, respectively.
It is also found that close to $P_{CO}^{c2}$,
when the propagation of the $O$ profile ceases, the velocity of the
$CO$ profile undergoes a sharp change. This behaviour can be
correlated with the first-order IPT between the stationary
reactive regime and the $CO$-poisoned state observed in the
ZGB model at $P_{2CO}$ (see figure 1).

So far, the main conclusions that can be drawn from figure 12(b)
can be summarized as follows: a) there are two critical pressures,
$P_{CO}^{c1}(L)$ and $P_{CO}^{c2}(L)$, which  depend on the
width of the channel, at which propagation of one profile or
the other stops; b) within these critical values, propagating
$CO$ and $O$ profiles coexist; c) $O$ profiles propagate faster
than $CO$ profiles.
All these observations appear in qualitative agreement with the
experimental results shown in figure 12(a) \cite{xx1}.
However, the underlying physics is different: in the simulations the
displacement of a poisoned phase by the invading reactive phase takes
place within a range of pressures where
the latter is unstable, while the former is stable. In contrast, the
experiment may show the propagation of coexisting phases within a
bistable regime \cite{xx1}.

So far, all those simulations of the ZGB model discussed above do
not attempt to describe the occurrence of oscillations in the
concentration of the reactants and in the rate of production,
which are well documented by numerous experiments
\cite{ron,ronert,bloc,bloc1}. In fact, it is well known that the catalytic
oxidation of $CO$ on certain $Pt$ surfaces exhibits oscillatory behaviour,
within a restricted range of pressures and temperatures, which is
associated with adsorbate-induced surface phase transitions \cite{ron,ronert}.
Since the aim of this paper is to describe the irreversible
critical behaviour of the reaction, the oscillatory behaviour will not
be further discussed. Therefore, the interested reader is addressed to 
recent developments involving the study of numerous lattice-gas models
aimed to explain the oscillations observed
experimentally \cite{nl,aren,evla,evpre,evjcp,kuzo,molle}.

Since the ZGB lattice gas reaction model is an oversimplified
approach to the actual processes involved in the catalytic
oxidation of CO, several attempts have been made in order to give
a more realistic description. Some of the additional mechanisms
and modifications added to the original model are the following:
(i) The inclusion of $CO$ desorption \cite{bloc,bloc1,kau,ziff3,alb1} that
causes the first-order IPT to become reversible and slightly rounded,
in qualitative agreement with experiments (figures 9 and 10).
(ii) Energetic interactions between reactants adsorbed on the
catalyst surface have been considered by various
authors \cite{kau,luq,satul}. In general, due to these interactions
the IPT's become shifted, rounded and occasionally they are no longer
observed \cite{kau,luq,satul,pele}.
(iii) Studies on the influence of the fractal nature of the catalyst
are motivated by the fact that the surface of most solids at the 
molecular level must be considered as a microscopic fractal, such as 
the case of cataliysts made by tiny fractal matallic cluster dispersed
in a fractal support or in a discontinuous thin metal films.
The fractal surfaces have been modeled by means of random
fractals, such as percolating clusters, \cite{alb2,alb3,nie2},
diffusion limited aggregates \cite{nie3} and also deterministic fractals,
such as Sierpinsky carpets, \cite{nie1,alb4},
etc. \cite{fra1,fra2}. One of the main findings of all these studies
is that the first-order IPT becomes of second-order for dimensions
$D_{F} < 2$. Since in $d = 1$ dimensions the ZGB model does not
exhibit a reaction window, one may expect the existence of
a `critical' lower fractal dimension capable a sustaining a
reactive regime. This kind of study, of theoretical interest in
the field of critical phenomena, remains to be addressed.
(iv) Also, different kinds of adsorption mechanisms, such as hot-dimer 
adsorption \cite{hot7}, local versus random
adsorption \cite{tam,chi}, nonthermal mechanisms involving
the precursor adsorption and diffusion of $CO$ molecules
\cite{kaalb}, the presence of subsurface oxygen \cite{ppaa},
etc. have been investigated.
(v) The influence of surface diffusion has also
been addressed using different approaches \cite{kau,dif1,dif2,jimi}.
Particularly interesting is the hybrid lattice-gas mean-field treatment
developed by Evans \etal \cite{jimi2} for the study of surface reactions
with coexisting immobile and highly mobile reactants.
(vi) Considering the Eley-Rideal mechanism \cite{er1,er2} as
an additional step of the set of equations (\ref{adco}-\ref{reac}), namely
including the following path
\begin{equation}
CO(g) + O(a) \rightarrow CO_{2}(g) + S 
\label{ERZGB}
\end{equation}
\noindent poisoning of the surface by complete occupation by
$O$ species is no longer possible preventing the
observation of the second-order IPT.
(vii) The influencia of surface defects, which has also been studied, 
merits a more detailed discussion because,
in recent years, considerable attention has been drawn to
studies of surface reactions on substrates that include defects or some 
degrees of geometric heterogeneity, which is not described by 
any fractal structure, as in the case of item iii).  
Interest in these types of substrates is based on the fact that the 
experiments have shown that inert species capable to block adsorption
sites, such a sulfur, deposit on the catalyst surface during 
the exhaust of the combustion gases.
Also crystal defects that are formed during the production of the catalyst
result in blocked sites. Other inhomogeneities consist of crystallographic
perturbations but may equally well involve active foreign surface atoms. 
Recently Lorenz \etal \cite{lorenz} have performed Monte Carlo simulations
using three different types of defective sites.
Site-1, that adsorbed neither $O$ nor $CO$ and Site-2 (Site-3)
that adsorbed $O$ ($CO$) but no $CO$ ($O$).
They found that $CO$ islands form around each defect near
$P_{2CO}$ ($P_{CO}=0.524-0.526$). The average density of $CO$ 
decays as a power-law of the radial distance from the 
defect ($\rho_{CO} = kr^{-m}$, $m=1.90(2)$),
and the average cluster size also obeys a power-law with the distance 
to spinodal point ($\Delta P = P_{CO}^{S} - P_{CO}$) with exponent $0.73$.
When defects are randomly distributed, with density $\theta_d$,
$P_{2CO}$ decreases linearly according to 
$P_{2CO} = -0.307 \theta_d + 0.5261$.
This model has also been investigated in the site and 
pair mean-field approximations \cite{hoenicke}. The pair approximation 
exhibits the same behaviour that the Monte Carlo simulation. 
The size of the reactive windows decreases with $\theta_d$ and 
the abrupt transition at $P_{2CO}$ becomes continuous 
(the same behaviour have been reported in a related model \cite{valencia}). 
However, unlike the analytical results, in the Monte Carlo simulation
there is a critical concentration above which the transition at $P_{2CO}$
becomes  continuous ($\theta_d = 0.75$ in agreement with previous results 
\cite{hovi}). In conclusion, various models shown that the presence of defects 
on the catalytic surface causes the $CO$ poisoning transition to 
occur at lower $P_{CO}$ values than on homogeneous surfaces. Also,
beyond some critical concentration of defects, the first-order IPT of 
the ZGB model becomes second-order. The overall effect of inert 
sites is to reduce the production of $CO_2$.
Furthermore, these findings provide an alternative explanation for the
absence of a second-order IPT into a $O$-poisoned state
observed in the experiments of $CO$ oxidation (see figure 2).
 
\subsection{The Catalytic Reaction Between Nitrogen Monoxide and 
Carbon Monoxide.}
\label{nomasco}

The catalytic reduction of $NO$ with various agents, including
$CO$, $H_{2}$, $NH_{3}$, hydrocarbons, etc., has been extensively
studied on $Pt$ and $Rh$ surfaces \cite{ron,ronert}, 
which are the noble metals 
used in automotive catalytic converters, due to the key role 
played by $NO_{x}$ emission in air pollution \cite{pollut}.
Aside from the practical importance, the catalytic reduction of 
$NO$ also exhibits a rich variety of dynamic phenomena
including multistability and oscillatory behaviour \cite{ron,ronert}.
Within this context, the catalytic reaction between $NO$ and
$CO$ is the subject of this section.

The archetypal model used in most simulations has early been proposed 
by Yaldran and Khan (YK) \cite{ykmodel}. As in the case of the 
ZGB model \cite{zgb}, the YK model is also a lattice gas reaction system 
based on the Langmuir-Hinshelwood mechanism. The reaction steps are 
as follows
\begin{equation}
NO(g) + 2 S \rightarrow N(a) + O(a)
\label{noad}
\end{equation}
\begin{equation}
CO(g) + S \rightarrow CO(a)
\label{coad}
\end{equation}
\begin{equation}
CO(a) + O(a) \rightarrow CO_{2}(g) + 2 S
\label{reco}
\end{equation}
\begin{equation}
N(a) + N(a) \rightarrow  N_{2}(g) + 2 S
\label{ren}
\end{equation}
where $S$ represents an unoccupied site on the catalyst surface, $2S$
represents a nearest neighbor (NN) pair of such sites, $(g)$ indicates
a molecule in the gas phase and $(a)$ indicates an species adsorbed on the
catalyst. The reactions given by equations (\ref{reco}) and (\ref{ren})
are assumed to be instantaneous (infinity reaction rate limit) while the
limiting steps are the adsorption events given by equations (\ref{noad}) and
(\ref{coad}). The YK model is similar to the ZGB model for the
$CO + O_{2}$ reaction, except that the $O_{2}$ is replaced by $NO$, and
NN $N$ atoms, as well as NN $CO-O$ pairs, react. For further details
on the YK model see \cite{ykmodel,bziff,menga,mengb,adick,yass}.

Early simulations of the YK model have shown \cite{ykmodel,bziff} 
that a reactive window is observed on the hexagonal lattice
while such kind of window is absent on the square lattice 
\cite{ykmodel,bziff}, pointing out the relevance of
the coordination number for the reactivity.
Therefore, we will first discuss Monte Carlo simulations 
of the YK model performed on the hexagonal 
(triangular) lattice. Subsequently, results
obtained for variants of the YK model that also exhibit reaction windows
on the square lattice will be presented. 

The simulation procedure in the standard ensemble is as follows: 
let $P_{NO}$ and $P_{CO}$ be the relative
impingement rates for $NO$ and $CO$, respectively, which are taken to
be proportional to their partial pressures in the gas phase. Taking
$P_{CO}\,+P_{NO}\, =\,1$, such normalization implies that the YK model
has a single parameter that is usually taken to be $P_{CO}$. $CO$ and
$NO$ adsorption events are selected at random with probabilities
$P_{CO}$ and $1-P_{CO}$, respectively. Subsequently, an empty site
of the lattice is also selected at random. If the selected species is
$CO$, the adsorption on the empty site occurs according to
equation (\ref{coad}). If the selected molecule is $NO$, a NN site of
the previously selected one is also chosen at random, and if such site
is empty the adsorption event takes place according to equation
(\ref{noad}). Of course, if the NN chosen site is occupied the adsorption
trial is rejected. After each successful adsorption event all NN sites
of the adsorbed species are checked at random for the occurrence of
the reaction events described by equations (\ref{reco}) and (\ref{ren}).

During the simulations, the coverages with $CO$, $O$ and $N$ 
($\theta_{CO}$, $\theta_{O}$ and $\theta_{N}$, respectively) as well
as the rate of production of $CO_{2}$ and $N_{2}$
($R_{CO_{2}}$, $R_{N_{2}}$, respectively) are measured.
The phase diagram of the YK model, shown in figure 13, is similar to 
that of the ZGB model \cite{zgb} shown in figure 1. In fact, in both cases
second- and first-order IPT's are observed. However, in
contrast to the ZGB model where the absorbing (poisoned) states are unique,
in the case of the YK model such states are mixtures of
$O(a)+N(a)$ and $CO(a)+N(a)$ as follows from the observation of the 
left and right sides of the phase diagram, respectively (figure 13(a)).
\begin{figure}
\begin{center}
\epsfxsize=6cm
\epsfysize=6cm
\epsfbox{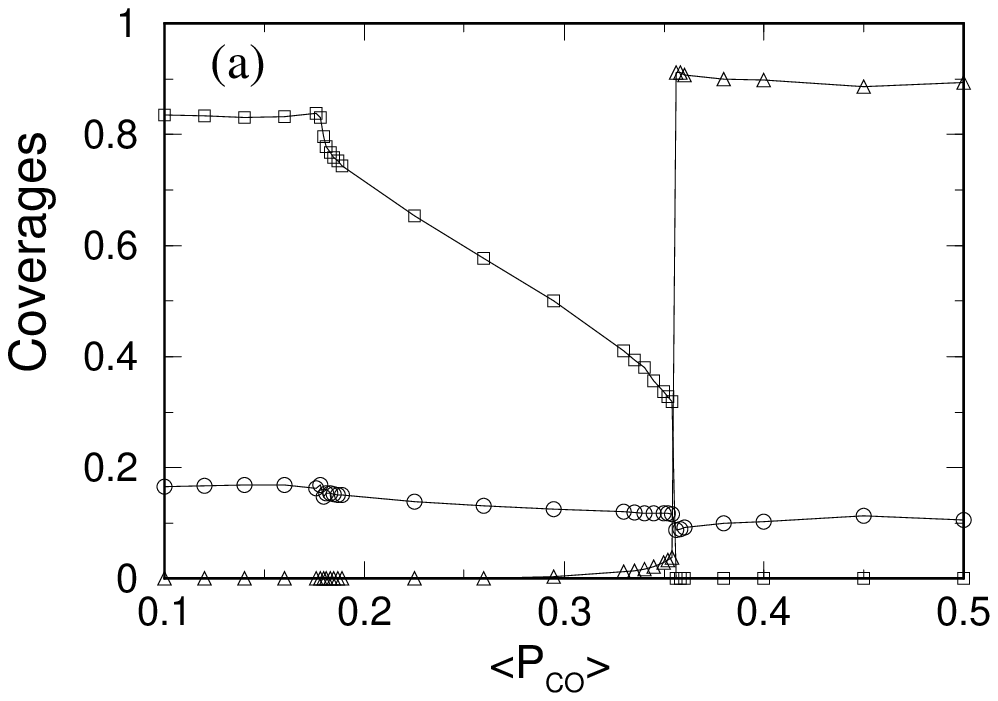}
\end{center}
\begin{center}
\epsfxsize=6cm
\epsfysize=6cm
\epsfbox{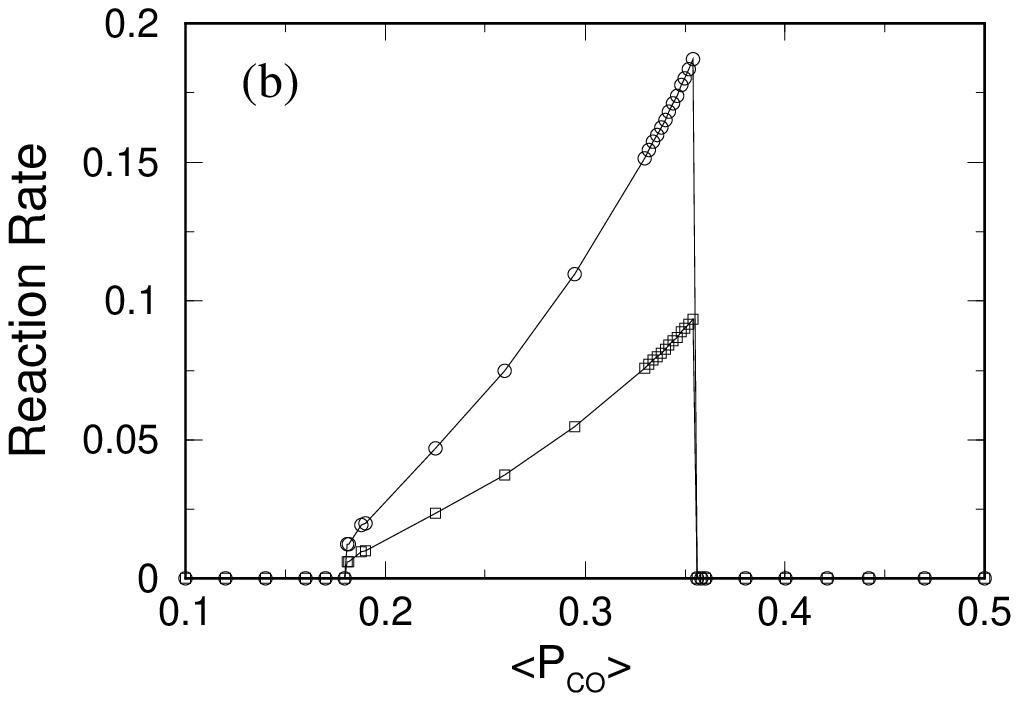}
\end{center}
\caption{Phase diagram of the YK model on the hexagonal lattice of size
L=128. (a) Plots of ${\theta}_{CO}(\bigtriangleup)$, 
${\theta}_{O}(\Box)$ and ${\theta}_{N}(\bigcirc)$ versus $<P_{CO}>$. 
(b) Plots of $R_{N_{2}}(\Box)$ and $R_{CO_{2}}(\bigcirc)$;
measured as units of number of $N_{2}$ and $CO_{2}$ molecules removed
from the lattice per unit area and time, respectively;
versus $<P_{CO}>$.}
\label{Figure13}
\end{figure}
The IPT observed close to $P_{1CO}=0.185 \pm 0.005$ \cite{ykmodel,bziff,yass}
is continuous and therefore of second-order (see figure 13).
More interesting, an abrupt first-order IPT is also observed close to  
$P_{2CO}=0.3545 \pm 0.005$ (figure 1(a) and (b)) \cite{ykmodel,bziff,yass}. 

Hysteretic effects close to the first-order IPT of the 
YK model have been investigated using the CC ensemble \cite{yass}
(see \Sref{nomasco}). For small lattices ($L \leq 64$) the
relaxation time is quite short, so that hysteretic effects are absent. 
This result is in agreement with similar measurements of the ZGB 
model (see figure 4(a)). On increasing the lattice size, 
hysteretic effects can be observed even for $L \geq 128$  
and they can unambiguously be identified for $L=256$, as shown
in figure 14(a). A vertical region located at
the center of the loop and slightly above $<\,P_{CO}\,>\,\approx\,0.35$,
as well as the upper spinodal point  $P^{US}_{CO}$, can easily 
be observed in figure 14. Furthermore, it is found that while the 
location of $P^{US}_{CO}$ is shifted systematically toward lower 
values when $L$ is increased, the location of the vertical
region (close to the center of the loops) remains almost fixed 
very close to $P_{CO}=0.3515$ \cite{yass}. Using lattices of 
size $L=1024$, the hysteretic effects are quite evident 
(see figure 14(b)) and also, the growing and decreasing branches 
of the loops are almost vertical. Also, the location of these
branches depends on the lattice size, as follows from the
comparison of figures 14(a) and (b).  
\begin{figure}
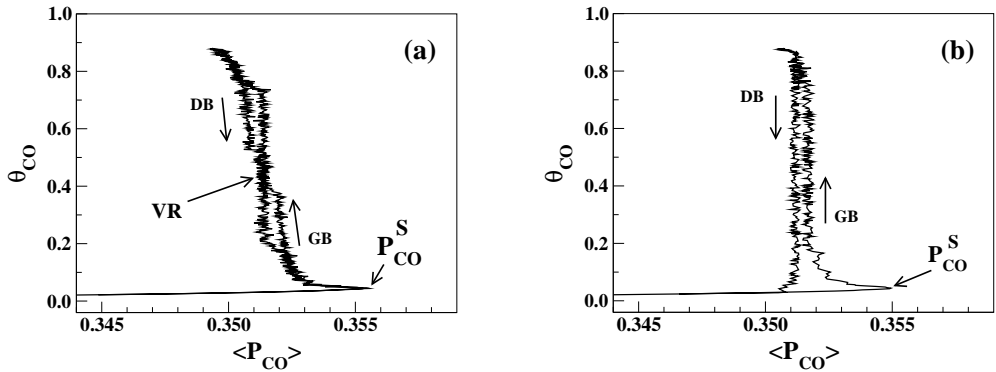

\begin{minipage}{7.0cm}
\begin{center}
\includegraphics[width=6.0cm]{Fig14a.eps}
\end{center}
\end{minipage}
\begin{minipage}{7.0cm}
\begin{center}
\includegraphics[width=6.0cm]{Fig14b.eps}
\end{center}
\end{minipage}
\caption {Plots of $\theta_{CO}$ versus $<P_{CO}>$ obtained using the 
CC ensemble and taking: (a) $L=256$ and (b) $L = 1024$. The arrows 
indicate the growing branch (GB), the decreasing branch (DB), the 
vertical region(VR) and the upper spinodal point ($P_{CO}^{S}$).
More details in the text.}
\label{Figure14}
\end{figure}
A more quantitative analysis on the behaviour of $P_{CO}$ corresponding to 
the different branches and the vertical region has also been
reported \cite{yass}. 
\begin{figure}
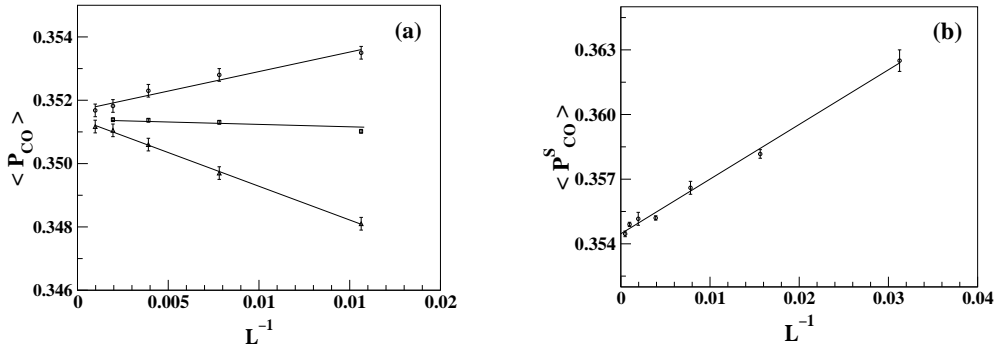

\begin{minipage}{7.0cm}
\begin{center}
\includegraphics[width=6.0cm]{Fig15a.eps}
\end{center}
\end{minipage}
\begin{minipage}{7.0cm}
\begin{center}
\includegraphics[width=6.0cm]{Fig15b.eps}
\end{center}
\end{minipage}
\caption {(a) Plots of $<P_{CO}>$ versus $L^{-1}$ measured in the 
growing branch ($\bigcirc$), decreasing branch ($\triangle$), 
and the vertical region ($\square$). 
The straight lines correspond to the best fits of the 
data that extrapolate to  $L\rightarrow\infty$.
(b) Plots of $P_{CO}^{S}$ versus $L^{-1}$. 
The straight line corresponds to the best fit of the data 
that extrapolates to $P_{CO}^{US}(L\rightarrow\infty)=0.3544(2)$.
More details in the text.}
\label{Figure15}
\end{figure}
Figure 15(a) shows the dependence 
of the location of the growing branch and the 
decreasing branch ($P^{GB}_{CO}$ and $P^{DB}_{CO}$, respectively) 
on the inverse of the lattice size. The $L$ dependence of $P_{CO}$ at 
the vertical region ($P^{VR}_{CO}$) is also shown for the sake of 
comparison. It has been found that the location of all relevant points, 
namely $P^{X}_{CO}$ with $X=GB,DB$ and $VR$, depends on the curvature 
radius ($s$) of the interface of the massive $CO$ cluster in contact 
with the reactive region. Such dependence can be written as follows
\begin{equation}
P^{X}_{CO}=P^{X}_{CO}(L\rightarrow\infty)+F^{X}(s) {,}
\label{ramas}
\end{equation}
where $P^{X}_{CO}(L\rightarrow\infty)$ is the location of the point under 
consideration after proper extrapolation to the thermodynamic limit and 
$F(s)$ is an $s$-dependent function. For the vertical region one has 
$s\rightarrow\infty$ and $P^{VR}_{CO}$ is almost independent of $L$, 
so $F^{VR}(\infty)\rightarrow\,0$, as shown in figure 15(a). In contrast, 
for the DB and the GB, $s$ is finite and of the order of $-1/L$ and $1/L$, 
respectively.  So, one has $F^{DB}(s)\,\approx\,-A/L$ while 
$F^{GB}(s)\,\approx\,B/L$, in agreement with the results shown in 
figure 15(a). The extrapolated points are
            
\begin{center}
$P^{GB}_{CO}(L\rightarrow\infty)=0.3514(3)$,
$P^{DB}_{CO}(L\rightarrow\infty)=0.3517(3)$ and
$P^{VR}_{CO}(L\rightarrow\infty)=0.35145(5)$
\end{center}
Also, $A\,\approx\,0.215(5)$ and 
$B,\approx\ 0.12(2)$ have been \cite{yass}.

On the basis of these results, $P^{VR}_{CO}(L\rightarrow\infty)$ 
has been identified as the coexistence point 
$P^{Coex}_{CO}\,\cong\,0.35145(5)$ in excellent agreement with an
independent measurement, $P_{2CO}=0.35140 \pm 0.00001)$, reported by
Brosilow and Ziff \cite{bziff}.
This result is in contrast with measurements performed with the ZGB model.
In fact, for the ZGB systems the vertical region is not observed 
while the locations of the growing and decreasing branches are 
almost independent of the lattice size (see figure 4). 
The explanation of the difference observed comparing both models, 
which  may be due to the different behaviour of the interface of 
the massive $CO$ cluster, is an interesting open question.

It has also been reported that the location of the upper spinodal 
point depends on the lattice size, as shown in figure 15(b).
This dependence of $P^{US}_{CO}(L)$ is due to local fluctuations 
in the $CO$ coverage that take place during the nucleation of 
the critical cluster \cite{yass}. 
The extrapolation of the data shown in figure 15(b) gives 
$P^{US}_{CO}(L\rightarrow\infty)\cong\,0.3544(2)$.  Furthermore, the 
coverage at this point is $\theta_{CO}^{US}\cong\,0.043(1)$. 
These results point out that in the thermodynamic limit the
spinodal point is very close to coexistence, i.e.,
$\Delta\,P_{CO}=P_{CO}^{US}-P_{CO}^{Coex}\,\cong\,0.003$. For the sake of
comparison it is worth mentioning that for the ZGB model one has   
$\Delta\,P_{CO}\,\cong\,0.0012$ (see \Sref{zgbsection}).

Further insight into the first-order IPT of the YK model can be gained 
performing  epidemic studies. However, in this case it is necessary to account 
for the fact that the poisoned (absorbing) state above coexistence 
is nonunique, since it is due to a mixture of $CO$ and $N$ atoms 
with coverage $\theta_{CO}\,\approx\,0.9$ and $\theta_{N}\,\approx\,0.1$, 
as shown in figure 13(a). So, the starting
configuration has to be obtained running the actual dynamics of the 
system slightly above coexistence until `natural' absorbing states 
suitable for the studies are generated.

Figure 16 shows results obtained performing epidemic simulations 
for various values of $P_{CO}$ including  $P_{CO}^{Coex}$,
$P_{CO}^{US}$, $P_{CO}^{DB}$, $P_{CO}^{GB}$
as well as a value close to coexistence but slightly inside the 
active region, namely $P_{CO}=0.347$.
\begin{figure}
\centerline{{\epsfysize=6.0 cm \epsffile{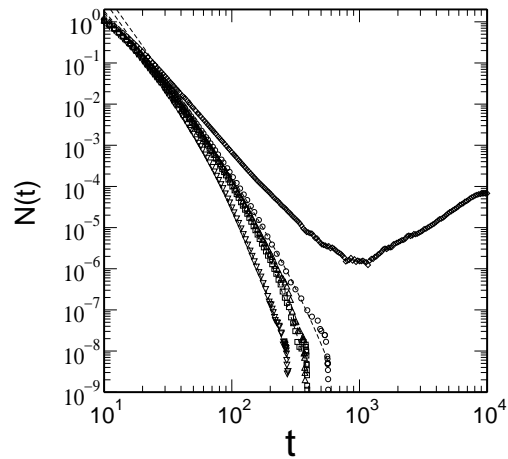}}}
\caption{ Log-log plots of the number of vacant sites $N(t)$ 
versus $t$ obtained performing epidemic simulations  
using lattices of size $L=256$. Results averaged up to $3$x$10^9$ different 
runs ($\triangledown$ $P^{S}_{CO}=0.3544$, 
$\square$ $P^{GB}_{CO}=0.3522$, $\triangle$ 
$P^{Coex}_{CO}=0.35145$, $\bigcirc$ $P^{DB}_{CO}=0.3506$ 
and $\lozenge$ $P_{CO}=0.3470$).}
\label{Figure16}
\end{figure}
From figure 16, it becomes evident that the 
method is quite sensitive to tiny changes of $P_{CO}$. The 
obtained curves are fitted by equation (\ref{anz}) with
$\eta_{eff} = -4.0 \pm 0.5$, indicating a markedly low 
survivability of the epidemic patches as compared with the ZGB 
model that gives $\eta_{eff} = -2.0 \pm 0.1$, as already discussed
in \Sref{zgbsection}. The main finding obtained using epidemic studies is 
that the occurrence of a power-law scaling behaviour close to 
coexistence can unambiguously be ruled out. This
result is in qualitative agreement with  data of
the ZGB model, see \Sref{zgbsection}. All these observations are also in 
agreement with the experience gained studying first-order reversible 
phase transitions where it is well established that correlations 
are short ranged, preventing the emergence of scale invariance.

It should be mentioned that several mean-field theories of the 
YK model have been proposed \cite{bziff,menga,adick,kor}.
Truncating the hierarchy of equations governing the cluster 
probabilities at the $1-$site level, a reasonable estimate
of the coexistence point given by $P_{2CO} = 0.3877$ is obtained 
\cite{bziff,menga} on the triangular lattice.
However, this approach fails to predict the second-order IPT
observed in the simulations (see figure 13) \cite{bziff,menga}.
Also, Kortl\"uke \etal \cite{kor} have derived an accurate
prediction of the second-order critical point at 
$P_{1CO} = 0.152$ using a two-site cluster approximation.
The prediction of this approach for the coexistence
point is less satisfactory, namely  $P_{2CO} = 0.393$. 
On the other hand, very recently an elaborated mean-field theory 
of the YK model has been developed up to the pair-approximation
level \cite{adick} that yields a very accurate estimation
of the coexistence point, namely  $P_{2CO} = 0.363$.  

As already mentioned above, the behaviour of the YK model
on the square lattice is radically different than that
observed on the triangular lattice. In fact, in the former
{\bf no} reactive stationary state has been 
observed \cite{ykmodel,bziff}, as shown in figure 17(a).
\begin{figure}
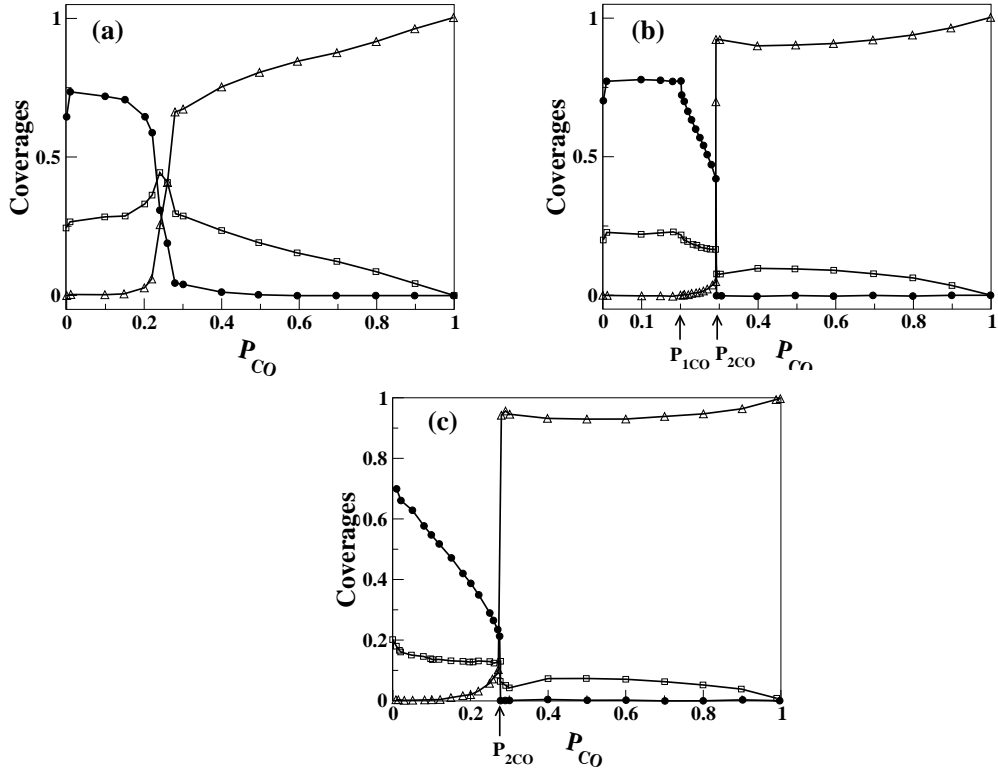

\begin{minipage}{7.0cm}
\begin{center}
\includegraphics[width=6.0cm]{Fig17a.eps}
\end{center}
\end{minipage}
\begin{minipage}{7.0cm}
\begin{center}
\includegraphics[width=6.0cm]{Fig17b.eps}
\end{center}
\end{minipage}
\begin{center}
\epsfxsize=6cm
\epsfysize=5cm
\epsfbox{Fig17c.eps}
\end{center}
\caption{Monte Carlo simulation results of the YK model on the 
square lattice showing plots of species coverages versus 
$P_{CO}$. $\theta_{O}$ (empty triangles), $\theta_{N}$ (empty squares),
and $\theta_{CO}$ (solid circles).
(a) Results obtained neglecting $N$ diffusion showing 
the absence of a reaction window such that the catalyst
always remains poisoned by mixtures of adsorbed species. 
(b) Results obtained considering $N$ diffusion. In this case the
YK model exhibits a reaction window. Adapted from reference \cite{pak}.
(c) Results obtained considering the influence 
of the Eley-Rideal (ER) mechanism and the diffusion on $N$. 
Adapted from reference \cite{paku}. More details in the text.}
\label{Figure17}
\end{figure}
However, simulation results
have shown that diffusion of $N$ (but not of $O$, or $CO$)
restores the possibility of a reactive state \cite{kor,pak},
as shown in figure 17(b). In fact, in this case a second-order
IPT is observed close to $P_{1CO} = 0.203 \pm 0.001$, while
a first-order IPT is found at $P_{2CO} = 0.293 \pm 0.001$ \cite{pak}.  
Also, Meng \etal \cite{mengb} have shown that by adding a new
reaction channel to YK model (equations (\ref{noad}-\ref{ren})), 
such that
\begin{equation}
CO(a) + N(a) \rightarrow CON(g) + 2 S
\label{addchan}
\end{equation}
the reactivity of the system becomes enhanced and consequently
a reaction window is observed on the square lattice.
This window exhibits a second-order IPT close to $P_{1CO} = 0.262$
and a first-order IPT close to $P_{2CO}= 0.501$. This behaviour
is reminiscent of that observed modeling the ZGB model, as 
discussed above.
   
On the other hand, assuming that the dissociation of $NO$ given by
equation (\ref{noad}) is preceded by a molecular adsorption on a
singe site, namely
\begin{equation}
NO(g) + S \rightarrow NO(a) ,
\label{noadmole}
\end{equation}
and
\begin{equation}
NO(a) + S \rightarrow N(a) + O(a),
\label{noaddiso}
\end{equation}
the YK model also exhibits a reaction window in the 
square lattice provided that both $NO$ and $CO$ desorption
are considered \cite{mengb}.     

Very recently, Khan \etal \cite{paku} have studied the influence 
of the Eley-Rideal (ER) mechanism (reaction of $CO$ molecules with already
chemisorbed oxygen atoms to produce desorbing $CO_{2}$) on the 
YK model on the square lattice. In the absence of $N$ diffusion, the 
added ER mechanism causes the onset of a reactive regime at extremely
low $CO$ pressures, i.e., for $P_{CO} \leq 0.03$. However, considering
the diffusion of $N$ species, the window becomes considerably 
wider and the reactive regime is observed up 
to $P_{2CO} \simeq 0.29$ where a first-order ITP is found \cite{paku},
as shown in figure 17(c).
This finding suggests that the incorporation of the ER mechanisms
does not affect the first-order IPT (see figure 17(b)). In contrast,
the second-order IPT is no longer observed as shown in figure 17(c).  

As in the case of the ZGB model, the bistable behaviour of the
YK model close to coexistence provides the conditions for the
displacement of reactive fronts or chemical waves. Within this context,
Tammaro and Evans \cite{tamma} have studied the reactive removal 
of unstable mixed $CO + NO$ layers adsorbed on the lattice.  
Furthermore, in order to account for the diffusion of the 
reactants, the hopping of all adsorbed species (except for $O$
atoms whose mobility is negligible) has been considered.   
Simulations are started with the surface fully covered by a mixture
$CO + NO$. This mixture is unstable since the vacation of a single site may
produce the dissociation of $NO$ (equation (\ref{noaddiso})) and its
subsequent reaction with $CO$ followed by desorption of the 
products and the generation of empty sites capable of
triggering the autocatalytic reaction. Due to the high mobility of most 
adsorbed species,  initially an exponential increase in the number of
highly dispersed vacancies is observed. Thereafter, a 
reaction front forms and propagates across the surface at 
constant velocity \cite{tamma}. It is also interesting to remark
that all simulation results are confirmed by an elaborated mean-field
treatment of chemical diffusion on mixed layers, incorporating
its coverage-dependent and tensorial nature, both of these features
reflecting the interference of chemical diffusion of adsorbed species
on surface by coadsorbed species \cite{tamma}. 
 
\subsection{Brief Overview of Other Surface Reaction Processes}
\label{bosta}

In addition to the ZGB and the YK models, numerous lattice gas
reaction models have also been proposed attempting to describe
catalyzed reaction processes of practical and academic interest.
Among others, the dimer-dimer (DD) surface reaction scheme of
the type $\frac{1}{2} O_{2} + H_{2} \rightarrow H_{2}O$
has been proposed in order to describe the catalytic oxidation of
hydrogen \cite{dd1}. Monte Carlo simulations of the DD model have
shown the existence of second-order IPT's and a rich variety
of irreversible critical behaviour \cite{dd1,dd2,dd3,dd4,yun,pako,puka}.
Relevant numerical results have also been qualitatively reproduced
by mean-field  calculations \cite{dd3}.

On the other hand, the catalytic synthesis of ammonia
from hydrogen and nitrogen on iron surfaces
($ N_{2} + 3 H_{2} \rightarrow 2 NH_{3}$) is among the
catalyzed reactions of major economical importance.
Ever since its discovery and technical realization,
the reaction has become the focus of fundamental
investigations. Very recently, various lattice gas
reaction models have been proposed and studied by means of
numerical simulations \cite{pakm,paev}.
The existence of IPT's has been observed,
as shown in figure 18 for the case of the model
proposed by Khan and Ahmad (KA) \cite{pakm}. Here the
Langmuir-Hinshelwood reaction mechanism is assumed and
the control parameter is the partial pressure of $H_{2}$.
As follows from figure 18, the KA model exhibits a
second-order (first-order) IPT close
to $P_{1H_2} \simeq 0.445$ ($P_{2H_2} \simeq0.585$),
resembling the behaviour of both the ZGB and the YK models
(see figures 1 and 13, respectively).

\begin{figure}
\begin{center}
\includegraphics[width=6.0cm]{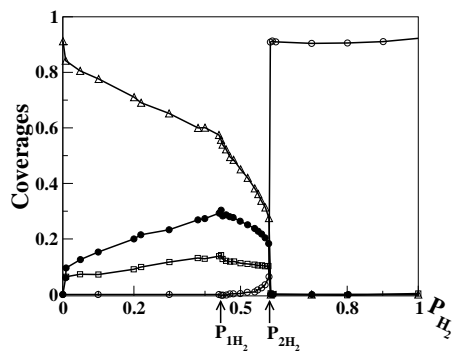}
\end{center}
\caption{Monte Carlo simulation results of the KA model on the 
square lattice showing plots of species coverages versus 
$P_{H_2}$. $\theta_{H}$ (open circles), $\theta_{N}$ (empty triangle)
$\theta_{NH}$ (open squares) and $\theta_{NH_{2}}$ (solid circles).
Adapted from reference \cite{pakm}.}
\label{Figure18}
\end{figure}

Recently, the simulation of reaction processes involving more than two
reactants has received growing attention. One example of these processes is
a multiple-reaction surface reaction model based upon both the ZGB and the
DD models \cite{mr1,mr2,mr3}. So, this ZGB-DD model may be applied to the
oxidation of CO in the presence of $H_2$-traces as well as to the oxidation
of hydrogen in the presence of CO-traces. Interest on this model is due to
various reasons. For example, the oxidation of hydrogen and carbon monoxide
plays a key role in the understanding of the process of hydrocarbon
oxidation. In fact, the oxy-hydrogen reaction mechanism contains the
chain-branching steps producing O, H and OH-radicals that attack 
hydrocarbon species. Also, CO is the primary product of hydrocarbon
oxidation and it is converted to carbon dioxide in a subsequent slow
secondary reaction. Furthermore, the ZGB-DD model exhibits interesting
irreversible critical behaviour with nonunique multi-component poisoned
states \cite{mr1,mr2,mr3}.

There are also models that are not aimed to describe any specific 
reaction system but, instead, they are intended to mimic generic 
reactions. Typical examples are the monomer-monomer model \cite{zgb},
the dimer-trimer model \cite{kova1,kova2,paakk},
the monomer-trimer model \cite{KAI}, etc. (see also
references \cite{rev1,rev4,rev5}.

On the other hand, in the literature there is a vast variety of lattice gas
reaction models following the spirit of the reaction described
in the above subsections. They all exhibit the same type of
irreversible critical behaviour at the transition, which is determined by a
common feature- the existence of an absorbing or poisoned state,
i.e., a configuration that the system can reach but from where
it cannot escape anymore, as discussed in detail in Section 1.2.
As already discussed, the essential physics for the occurrence
of IPT's is the competition between proliferation and death of
a relevant quantity. So, it is not surprising that a large number
of models satisfying this condition, and aimed to describe quite
diverse physical situations, have been proposed and studied.
Some examples of models exhibiting second-order IPT's are,
among others `directed percolation' as a model
for the systematic dripping of a fluid through a lattice
with randomly occupied bonds \cite{rev1,dp1,dp2},
the `contact process' as a simple lattice model for the spreading
of an epidemics \cite{rev1,rev2,har,dur},
`autocatalytic reaction-diffusion models' aimed to describe the
production of some chemical species \cite{sch},
`the stochastic game of life' as a model for a
society of individuals \cite{mon1,mon2},
`forest fire models' \cite{clar,evaff}, `branching annihilating random walkers'
with odd number of offsprings \cite{baw1,baw11,baw12}, 
epidemic spreading without 
immunization \cite{epwi}, prey-predator systems \cite{ale},
the Domany-Kinzel cellular automata \cite{dk2}, etc.
For an excellent review on this subject see \cite{haye}.

The common feature among all these models is that they
exhibit second-order IPT's belonging to the universality
class of directed percolation (DP), the Langevin
equation (\Sref{lang}) being the corresponding
field-theoretical representation. The robustness of these
DP models with respect to changes in the microscopic dynamic
rules is likely their most interesting property. Such
robustness has led Janssen and Grassberger \cite{dpcon1,dpcon2}
to propose the so-called DP conjecture, stating that models
in the DP universality class must satisfy the following conditions:
(i) They must undergo a second-order IPT from a fluctuating
active state to a unique absorbing state. (ii) The transition
has to be characterized by a positive single-component
order parameter. (iii) Only short-range processes are allowed
in the formulation of the microscopic dynamic rules.
(iv) The system has neither additional symmetries
nor quenched randomness. In spite of the fact that
the DP conjecture has not been proved rigorously,
there is compelling numerical evidence supporting it \cite{note99}.

So far, DP appears to be the generic universality
class for IPT's into absorbing states, having
a status similar to their equilibrium counterpart,
namely the venerated Ising model. However, despite
the successful theoretical description of the DP
process, there are still no experiments where the
critical behaviour of DP has been observed.
Therefore, this is a crucial open problem in the
field of IPT's. For further discussions see for instance
\cite{dpopen}.

\section{Conclusions}

The study of irreversible critical behaviour in reaction systems
has attracted the attention of many physicists and
physical-chemists for more than four decades. On the one hand,
second-order IPT's are quite appealing since, like the archetypal
case of DP, they are observed in simple models in terms of
their dynamic rules. Nevertheless, second-order behaviour
is highly nontrivial and has not yet been solved exactly,
even using minimal models in one dimension. 
Furthermore, the critical exponents
are not yet known exactly. Field-theoretical calculations
and numerical simulations have greatly contributed to the
understanding of second-order irreversible behaviour.
Most systems lie in the universality class of DP, which
plays the role of a standard  universality class
similar to the Ising model in equilibrium statistical physics,
and the reason for few of the exceptions found are very well
understood. The main challenges in the field are, from the
theoretical point of view, the achievement of exact solutions
even for simple models and, from the point of view of the
experimentalists, the realization of key experiments
unambiguously showing DP behaviour.

On the other hand, the scenario for the state of the art in
the study of first-order IPT's is quite different. Firstly,
there is stimulating experimental evidence of the existence
of abrupt (almost irreversible) transitions, while hysteretic
effects and bistable behaviour resembling first-order
like behaviour have been observed in numerous catalyzed reaction
experiments. Secondly, one still lacks a theoretical
framework capable describing first-order
IPT's and theoretical efforts are being addressed to
the development of mean-field approaches with different
degrees of sophistication.

The achievement of a theoretical framework enabling the treatment
of irreversible critical behaviour and the gathering of further
experimental evidence, including the accurate measurement of
critical exponents, are topics of high priority that will
certainly contribute to the development of a general theory
of the physics of far-from equilibrium processes.

\ack This work is financially supported by CONICET, UNLP and 
ANPCyT (Argentina). We are grateful with numerous colleagues 
for stimulating discussions.


\begin{thebibliography}{}
\bibitem{note99} Only few exceptions of DP are known so far.
However, they all violate at least one of the essential
conditions required for the validity of the DP conjecture,
as dicussed in \Sref{bosta}. For some examples see reference
\cite{haye}.
\end{thebibliography}
\end{document}